\documentclass[aps,prx,showpacs,floatfix,twocolumn,superscriptaddress,longbibliography]{revtex4-2}

\usepackage[T1]{fontenc}
\usepackage{textcomp}
\usepackage{graphicx} 
\usepackage{latexsym}
\usepackage{amsmath,amssymb}
\usepackage{bm} 
\usepackage{xcolor}
\usepackage{accents}
\usepackage{braket}
\usepackage{float}
\usepackage{verbatim}
\usepackage{soul}
\usepackage{glossaries}
\usepackage{sidecap}
\usepackage{hyperref}
\hypersetup{
    colorlinks=true,
    linkcolor=blue,
    filecolor=magenta,
    urlcolor=red,
    citecolor=blue,
}

\newacronym{CO}{CO}{charge order}
\newacronym{JDOS}{JDOS}{joint density of states}
\newacronym{BS}{BS}{bond-stretching}
\newacronym{RIXS}{RIXS}{resonant inelastic x-ray scattering}
\newacronym{REXS}{REXS}{resonant elastic x-ray scattering}
\newacronym{XAS}{XAS}{x-ray absorption spectrum}
\newacronym{EELS}{EELS}{electron energy loss spectroscopy}
\newacronym{EPC}{EPC}{electron-phonon coupling}
\newacronym{CDW}{CDW}{charge density wave}
\newacronym{SDW}{SDW}{spin density wave}
\newacronym{FWHM}{FWHM}{full-width at half-maximum}
\newacronym{INS}{INS}{inelastic neutron scattering}
\newacronym{DFT}{DFT}{density functional theory}
\newacronym{GGA}{GGA}{generalized gradient approximation}
\newacronym{UHB}{UHB}{upper Hubbard band}
\newacronym{ZSA}{ZSA}{Zaanen-Sawatzky-Allen}
\newacronym{ZRS}{ZRS}{Zhang-Rice singlet}
\newacronym{ED}{ED}{exact diagonalization}
\newacronym{CEF}{CEF}{crystal electric field}
\newacronym{2D}{2D}{two-dimensional}
\newacronym{TM}{TM}{transition-metal}
\newacronym{LDA}{LDA}{local density approximation}
\newacronym{DMFT}{DMFT}{dynamical mean field theory}
\newacronym{NSLS-II}{NSLS-II}{National Synchrotron Light Source-II}
\newacronym{SOC}{SOC}{spin-orbit coupling}

\begin{document}

\title{Witnessing Quantum Entanglement Using Resonant Inelastic X-ray Scattering}

\author{Tianhao Ren}
\affiliation{Condensed Matter Physics and Materials Science Division, Brookhaven National Laboratory, Upton, New York 11973, USA}

\author{Yao Shen}
\email{yshen@iphy.ac.cn, Present address: Beijing National Laboratory for Condensed Matter Physics, Institute of Physics, Chinese Academy of Sciences, Beijing 100190, China; Institute of Physics, Chinese Academy of Sciences, Beijing 100190, China}
\affiliation{Condensed Matter Physics and Materials Science Division, Brookhaven National Laboratory, Upton, New York 11973, USA}

\author{Marton Lajer}
\email{mlajer@bnl.gov}
\affiliation{Condensed Matter Physics and Materials Science Division, Brookhaven National Laboratory, Upton, New York 11973, USA}

\author{Sophia F. R. TenHuisen}
\affiliation{Department of Physics, Harvard University, Cambridge, Massachusetts 02138, USA}

\author{Jennifer Sears}
\author{Wei He}
\affiliation{Condensed Matter Physics and Materials Science Division, Brookhaven National Laboratory, Upton, New York 11973, USA}

\author{Mary H. Upton}
\author{Diego Casa}
\affiliation{Advanced Photon Source, Argonne National Laboratory, Argonne, Illinois 60439, USA}

\author{Petra Becker}
\affiliation{Section Crystallography, Institute of Geology and Mineralogy, University of Cologne, 50939 K\"{o}ln, Germany}

\author{Matteo Mitrano}
\affiliation{Department of Physics, Harvard University, Cambridge, Massachusetts 02138, USA}

\author{Mark P. M. Dean}
\email{mdean@bnl.gov}
\affiliation{Condensed Matter Physics and Materials Science Division, Brookhaven National Laboratory, Upton, New York 11973, USA}
\author{Robert M. Konik}
\email{rmk@bnl.gov}
\affiliation{Condensed Matter Physics and Materials Science Division, Brookhaven National Laboratory, Upton, New York 11973, USA}

\date{\today}

\begin{abstract}
Although entanglement is both a central ingredient in our understanding of quantum many-body systems and an essential resource for quantum technologies, we only have a limited ability to quantify entanglement in real quantum materials. Thus far, entanglement metrology in quantum materials has been limited to measurements involving Hermitian operators, such as the detection of spin entanglement using inelastic neutron scattering. Here, we devise a method to extract the quantum Fisher information (QFI) from non-Hermitian operators and formulate an entanglement witness for resonant inelastic x-ray scattering (RIXS). Our approach is then applied to the model iridate dimer system Ba$_3$CeIr$_2$O$_9$ and used to directly test for entanglement of the electronic orbitals between neighboring Ir sites. We find the entanglement can be detected if we account for the expected symmetries, parity, and electron number conservation, of the dimer system.  We also consider the roles that the incident and outgoing x-ray polarizations and the incident photon energy play in entanglement detection. Our protocol provides a new handle for entanglement detection in quantum materials.

\end{abstract}

\maketitle

\section*{Introduction}

\begin{figure*}[t]
	\centering
	\includegraphics[width=\textwidth]{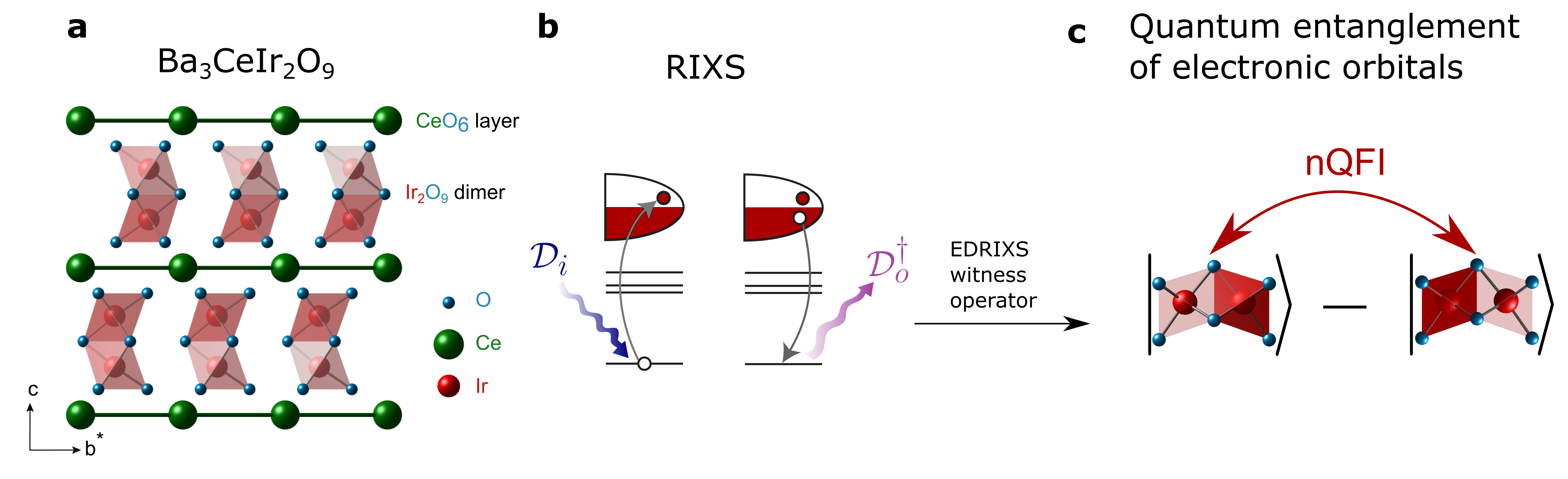}
	\caption{\textbf{Witnessing orbital entanglement using RIXS}. \textbf{a} The crystal structure of Ba$_3$CeIr$_2$O$_9$ is shown with the dimer units highlighted in red, which host Ir $5d$ orbitals that are separated by distance $d$. \textbf{b} In this letter, we develop a joint experimental-theoretical method to obtain an orbital entanglement witness from the RIXS intensity, which probes the material through a scattering process involving the absorption (${\cal D}_i$) and the subsequent emission (${\cal D}^\dagger_o$) of x-rays shown as wavy blue and purple lines. \textbf{c)} We first obtain the witness operator from the RIXS intensity using the numerical codebase EDRIXS and subsequently obtain the normalized quantum Fisher information (nQFI) for detection of the quantum entanglement.}
	\label{fig:overview}
\end{figure*}

Multipartite quantum entanglement refers to the entanglement of a quantum system across multiples of its subsystems.  While all entanglement measures express in some form the non-locality that is a fundamental aspect of quantum mechanics, the phenomenology of multipartite entanglement is, in general, much richer than that given by bipartite measures such as the R\'enyi and von Neumann entanglement entropies.   Multipartite entanglement is also a driver of quantum technologies where entanglement across multiple sites acts as a resource \cite{RevModPhys.91.025001} that enables quantum communication \cite{Piveteau_2022}, enhanced quantum metrology \cite{Giovannetti2011}, quantum sensing \cite{RevModPhys.89.035002}, quantum machine learning \cite{garcia2022systematic}, and quantum imaging \cite{Magana-Loaiza_2019,Moreau2019}. Given its centrality, it is important to be able to characterize multipartite entanglement in materials.  

Although there are many approaches for quantifying bipartite entanglement in synthetic quantum systems \cite{Blatt_2005,Jurcevic_2014,PhysRevA.79.042334,Oberthaler_2008,Oberthaler_2010,PhysRevLett.115.035302,Cramer_2013,PhysRevLett.106.150404,PhysRevLett.109.020504,PhysRevLett.109.020505,PhysRevA.98.052334,doi:10.1126/science.aau4963,Pitsios_2017}, most of these are impractical for solid-state quantum materials, due to either the large number of atoms in such materials or the lack of fine-scale control. Entanglement in magnetic materials between spin pairs can be inferred from magnetic susceptibility \cite{Ghosh_2003,PhysRevA.73.012110,PhysRevB.75.054422,Das_2013,Sahling_2015}, magnetic specific heat \cite{Singh_2013}, or the dynamical spin structure factor \cite{PhysRevLett.99.087204,PhysRevB.73.134404,doi:10.1073/pnas.0703293104}.  A more general approach both capable of detecting arbitrary amounts of multipartite entanglement and that is potentially compatible with a variegated set of experimental solid state probes is the quantum Fisher information (QFI) \cite{Tennant_PhysRevB.103.224434}. The QFI is a concept from quantum metrology \cite{helstrom1969quantum,PhysRevLett.72.3439,Giovannetti2011,Toth_2014} that pertains to the probability distribution of measurements and the corresponding parameter estimation in a multiparticle quantum system. If the precision of a parameter estimation exceeds the classical limit, then it can be deduced that the system must have multipartite entanglement \cite{PhysRevLett.102.100401,pezze2014quantum,hauke2016, QFIMPE1,PhysRevB.99.045117}. As shown in Ref.~\cite{hauke2016}, the QFI can be deduced from appropriately weighted energy-integrals of dynamical susceptibilities. This approach depends on the operator associated with the susceptibility being both Hermitian and having bosonic statistics. Since the spin operator fulfills this criterion, the formulae in Ref.~\cite{hauke2016} can be directly applied to inelastic neutron scattering, and this has been successfully used to detect entanglement in quasi-one-dimensional quantum magnets \cite{PhysRevResearch.2.043329,Tennant_PhysRevB.103.224434,PhysRevLett.127.037201}. 

RIXS is a fast-evolving experimental technique that can probe charge, spin, and orbital degrees of freedom \cite{RevModPhys.83.705, RevModPhys.93.035001,DEAN20153,Mitrano2024exploring}. Given its flexibility to probe multiple degrees of freedom and its ability to measure very small sample volumes and in ultrafast pump-probe modalities, it offers exciting possibilities to expand the frontier of entanglement metrology in quantum materials \cite{Dean2012spin, Dean2016ultrafast, Mitrano2024exploring}. However, while the RIXS intensity is bosonic, it is non-Hermitian, and thus Ref.~\cite{hauke2016}'s formalism cannot be directly applied.  One theoretical suggestion to circumvent this difficulty is to convert the RIXS intensity into an approximate estimate for the dynamical spin structure factor \cite{neutron_from_trRIXS}. 

Here instead we take a different approach.  The key result of this letter is that we show how to extend Ref.~\cite{hauke2016}'s formalism to cover non-Hermitian operators and use it to directly formulate an entanglement witness that exploits the full complexity of the RIXS response. We use this theoretical advance to convert measured RIXS intensities of Ba$_3$CeIr$_2$O$_9$ to the normalized QFI. This material features face-sharing Ir octahedra, which makes it an ideal test case for detecting a prototypical two-partite entanglement between its $t_{2g}$ electronic orbitals.  Iridium based materials containing face-sharing motifs, have attracted considerable interest as these interacting orbitals can realize cancellation of local moment magnetism, spinons, and quantum spin liquids under different circumstances
\cite{Kugel2015Spin, Streltsov2016covalent, Wang2019direct, Cao2020quantum, Shen2022emergence}. An overview of this process is illustrated in Fig.~\ref{fig:overview}.

\section*{Designing RIXS as an Entanglement Witness}
\textit{Entanglement:}
Quantum entanglement is a property of the many-body wave function of a quantum system. To define multipartite entanglement \cite{bengtsson2016brief,walter2017multipartite}, first consider a pure state $\rho=|\psi\rangle\langle \psi|$ of a $N$-site system. The state is said to be $m$-separable if one can write the state as a product of states $\rho_j$ involving $m_j\leqslant m$ sites:
\begin{equation}
    \rho=\otimes_{j=1}^M\rho_j, \quad \sum_{j=1}^Mm_j=N,
\end{equation}
where $M$ is the number of independent subsystems (or blocks) into which $|\psi\rangle$ factorized.

If the state cannot be further factorized into smaller pieces, it is said to possess $m$-partite entanglement. A mixed, i.e., thermal, state, $\rho=\sum_i\rho_i|\psi_i\rangle\langle\psi_i|$, is said to possess $m$-partite entanglement if the maximally entangled wavefunction, $|\psi_i\rangle$, in the mixture has $m$-partite entanglement \cite{PhysRevLett.102.100401,QFIMPE1,pezze2014quantum,Tennant_PhysRevB.103.224434}. The physical consequence of entanglement is that a measurement of one site in the system, represented by the action of an operator on that site, necessarily affects the other entangled sites. 

\textit{Quantum Fisher information:} The quantum Fisher information (QFI) provides a way to connect bounds on the multipartite entanglement with spectroscopic probes that can be measured experimentally.  The QFI, $F_Q(\rho,\hat{A})$, 
 governs the sensitivity of the quantum density matrix $\rho$ to unitary rotations $U^\dagger\rho U$, $U=e^{\mathrm{i}\theta\hat{A}}$, defined in terms of a Hermitian operator $\hat{A}$ \cite{metrology2,pezze2014quantum}. The variance by which the parameter $\theta$ can be determined by a single measurement is given by $(\Delta\theta)^2 \geq F_Q(\rho,\hat{A})^{-1}$.

The QFI is expressible in terms of the response function relative to the operator $\hat{A}$, the key observation of Ref.~\cite{hauke2016}:
\begin{eqnarray}
\label{eq:integral}
    F_Q(\rho, \hat{A}) 
     &=& 4 \int_{0}^{\infty} \mathrm{d} \omega \tanh \left(\frac{\beta \omega}{2}\right) \chi^{\prime \prime}_{\hat{A}\hat{A}}(\omega),
\end{eqnarray}
where $\chi''_{\hat{A}\hat{A}}(\omega)$ is the imaginary part of the retarded correlation function 
\begin{equation}
	 \chi''_{\hat{A}\hat{A}}(\omega)\!=\!{\rm Im}\bigg[i \int_0^{\infty} \!\frac{\mathrm{d}t}{\pi}~e^{\mathrm{i} \omega t} \operatorname{Tr}\left(\rho\left[\hat{A}(t), \hat{A}(0)\right]\right)\bigg].
\end{equation}
It has been shown that if the QFI exceeds a certain bound \cite{QFIMPE1,PhysRevLett.102.100401,PhysRevA.85.022322,doi:10.1126/science.1250147}, the state is guaranteed to have a certain level of multipartite entanglement.  In metrological terms, this means that the presence of multipartite entanglement makes measurements of the parameter, $\theta$, more accurate than would be possible if the state were purely classical without any entanglement. This bound is typically computed by assuming that the operator $\hat{A}$ is a sum over local operators defined at each site of the system $\hat{A}=\sum_{i=1}^N\hat{A}_i$.  If we denote the maximum and minimum eigenvalues of $\hat{A}_i$ as $a_{i,\text{max}}$ and $a_{i,\text{min}}$ respectively and if the QFI satisfies
\begin{equation}
\label{eq:nQFI}
    \frac{F_Q(\rho,\hat{A})}{\sum_{i=1}^N(\Delta a_i)^2}>m, \quad \Delta a_i=a_{i,\text{max}}-a_{i,\text{min}},
\end{equation}
then the state $\rho$ is guaranteed to have $(m+1)$-partite entanglement.  
The left-hand side of the inequality in Eq.~\eqref{eq:nQFI} is referred to as the normalized QFI (nQFI). 

Exploiting this connection between $F_Q(\rho, \hat{A})$ and $\chi''_{\hat{A}\hat{A}}$, the multipartite entanglement of quasi-1D quantum magnets was determined from neutron scattering measurements of the dynamic spin structure factor, i.e., $\hat{A}=\hat{S}$ \cite{PhysRevResearch.2.043329,Tennant_PhysRevB.103.224434,PhysRevLett.127.037201}. Unlike with neutron scattering, there are two key challenges to applying this approach to RIXS: i) the relevant operator $\hat{A}$ for RIXS is not Hermitian so Ref.~\cite{hauke2016}'s formalism cannot be straightforwardly employed; ii) the RIXS operator is not strictly speaking an operator that is a sum of operators defined on single sites. We now consider both challenges in turn.

\textit{RIXS as an entanglement witness:} In RIXS, x-rays with energy $\omega_{\text{in}}$, momentum $k_i$, and polarization $\boldsymbol{\epsilon}_i$ excite a core electron to the valence band. The resulting core hole is then refilled by a valence electron and an outgoing photon with energy loss $\omega$, momentum $k_o$ and polarization $\boldsymbol\epsilon_o$ is emitted. The corresponding RIXS intensity is described by the Kramers-Heisenberg formula \cite{RevModPhys.83.705, WANG2019151, Mitrano2024exploring}

\begin{eqnarray}\label{eq:RIXS}
I_{\mathrm{RIXS}}(\omega_{\mathrm{in}}, \omega, k_i, k_o, \boldsymbol\epsilon_i,\boldsymbol\epsilon_o) &=& \sum_{g,f}\frac{\Gamma / \pi}{(\omega-E_{f}+E_{g})^2+\Gamma^2} \times\cr\cr
&& \hskip -1.35in \frac{e^{-\beta E_{g}}}{Z} \bigg|\sum_{n,j}\frac{\langle f,k_o,\epsilon_o|\hat{\cal D}^\dagger_{jo}|n\rangle\langle n|\hat{\cal D}_{ji}|g,k_i,\epsilon_i\rangle}{\omega_{\text {in }}-E_n+E_{g}+i\Gamma_c}\bigg|^2 .
\end{eqnarray}
Here $\ket{g,k_i,\epsilon_i}$ and $\ket{f,k_o,\epsilon_o}$ mark the system's (material+photon) initial and final eigenstates. $E_{g}$ is the initial energy of the material absent a core hole governed by the Hamiltonian $H_0$.  $\ket{n}$, in contrast, is the eigenstate of the material's Hamiltonian, $H_n$, in the presence of a core hole with eigenvalue $E_n$. $\hat{\cal D}_{ji}$ is the dipole operator at site $j$ appropriate to the absorption of the incident photon with momentum $k_i$ and polarization $\boldsymbol\epsilon_i$ (in Sec.~I of \cite{supp}, we give in detail the form of the dipole operators). ${\cal D}^\dagger_{jo}$, governing photon emission, is similarly defined.  $T=(k_B\beta)^{-1}$ is the sample temperature and $Z$ is the corresponding partition function. $\Gamma$ is the inverse lifetime of the final state, while $\Gamma_c$ is the inverse lifetime of the core-hole state. For experiments that do not discriminate the final-state polarization, we also need to sum over $\boldsymbol\epsilon_o$ in Eq.~\eqref{eq:RIXS}. 

Inspecting Eq.~\eqref{eq:RIXS}, we can identify the RIXS operator $\hat{A}_{R}^{\dagger}(k_o,\boldsymbol\epsilon_o;k_i,\boldsymbol\epsilon_i)$, that scatters a photon $(k_i,\boldsymbol\epsilon_i)$ into a photon $(k_o,\boldsymbol\epsilon_o)$ as
\begin{eqnarray}
\label{eq:Melement}
    \hat{A}_{R}^{\dagger}(k_o,\boldsymbol\epsilon_o;k_i,\boldsymbol\epsilon_i)&=&\sum_{n\atop j=1,2}
     {\cal D}_j^\dagger(k_o,\boldsymbol\epsilon_o)|n\rangle\langle n| {\cal D}_j(k_i,\boldsymbol\epsilon_i)\cr\cr 
     && \hskip -.7in \frac{1}{\omega_{\rm in}+i\Gamma_c-E_n+H_0}\otimes |k_o,\boldsymbol\epsilon_o\rangle\langle k_i,\boldsymbol\epsilon_i|,
\end{eqnarray}
where the kets $|k_{i,o},\boldsymbol\epsilon_{i,o}\rangle$ mark the incoming and outgoing photon states.  The resulting matrix elements of this operator take the form 
\begin{eqnarray}
    \braket{f,k_o,\epsilon_o|\hat{A}_{R}^{\dagger}|g,k_i,\epsilon_i}&=&\cr\cr 
    &&\hskip -1.0in \sum_{n\atop j=1,2}
    \frac{\braket{f|{\cal D}_j^\dagger(k_o,\boldsymbol\epsilon_o)|n}\braket{n| {\cal D}_j(k_i,\boldsymbol\epsilon_i)|g} }{\omega_{\rm in}+i\Gamma_c-E_n+E_g} .
\end{eqnarray}
In the case of when we sum over outgoing polarizations the RIXS operator is given as 
\begin{equation}
\hat{A}_{R}^{\dagger}(k_o;k_i,\boldsymbol\epsilon_i) = \sum_{\boldsymbol\epsilon_o}\hat{A}_{R}^{\dagger}(k_o,\boldsymbol\epsilon_o;k_i,\boldsymbol\epsilon_i)
\end{equation}
In either case, the RIXS intensity can be written as
\begin{equation}
I_{\mathrm{RIXS}}\left(\omega\right)=\frac{\chi''_{\hat{A}_R{\hat{A}_R}^{\dagger}}(\omega)}{1-e^{-\beta \omega}},\label{RIXSvsresponse}
\end{equation}
where we have omitted the dependencies on the incoming and outgoing photons for notational conciseness. Relation \eqref{RIXSvsresponse} is exact in the $\Gamma\rightarrow0$ limit where the energy loss profile due to instrumental and final-state broadening can be replaced by a delta function. Here the thermal trace implicit in $\chi''_{\hat{A}_R\hat{A}^\dagger_R}$ is done with respect to $\rho = Z^{-1}\sum_g e^{-\beta E_g}|g\rangle\langle g|\otimes|k_i,\boldsymbol\epsilon_i\rangle\langle k_i,\boldsymbol\epsilon_i|$.

To convert $I_{\mathrm{RIXS}}(\omega)$ into a witness of entanglement despite the non-Hermiticity of $\hat{A}_R$, we consider instead the real and imaginary parts of the RIXS operator:
\begin{equation}
\label{eq:sep}
    \hat{A}_{R,\text{Re}}=\frac{1}{2}\left(\hat{A}_R+\hat{A}_R^{\dagger} \right), \quad \hat{A}_{R,\text{Im}}=\frac{1}{2i}\left(\hat{A}_R-\hat{A}_R^{\dagger} \right).
\end{equation}
Each of these operators are Hermitian.
We do note that this connects to a larger body of work on how to understand the QFI and, more broadly, quantum parameter estimation when an underlying quantum state is experiencing non-unitary evolution \cite{Yu2023quantum,e15093361}.  However as far as we understand, the concern there has not been on the detection of multipartite entanglement per se, the focus of our work.

With these Hermitian operators in hand, We can thus define the QFI for both:
\begin{equation}
F_Q(\rho, \hat{A}_{R,\text{Re}}),~~ F_Q(\rho, \hat{A}_{R,\text{Im}}).
\end{equation}
Each of these QFI's individually is not connected to a spectroscopy that can be experimentally measured.  However, as one of our key results, the sum of the two is related to the RIXS intensity $I_{\text{RIXS}}(\omega)$ measured at all frequencies, both positive and negative:
\begin{equation}
\label{RIXS_QFI}
    \begin{split}
        & F_Q^\mathrm{RIXS}\equiv F_Q(\rho,\hat{A}_{R,\text{Re}})+F_Q(\rho,\hat{A}_{R,\text{Im}})\\
        &=2\int_{0}^{\infty}\mathrm{d}\omega\tanh\left(\frac{\beta\omega}{2}\right) \left[ \chi''_{\hat{A}_R{\hat{A}_R}^{\dagger}}(\omega)+\chi''_{{\hat{A}_R}^{\dagger}\hat{A}_R}(\omega)\right] \\
        &=2\int_{-\infty}^{\infty}\mathrm{d}\omega\tanh\left(\frac{\beta\omega}{2}\right)\left(1-e^{-\beta\omega}\right)I_{\text{RIXS}}(\omega).
    \end{split}
\end{equation}
The derivation of eq. \eqref{RIXS_QFI} is discussed in Sec. II of \cite{supp}.

Thus in order to determine the QFI we need to compute the RIXS response from energy gain ($\omega<0$) processes.  While the intensity of these processes are thermally suppressed, the $(1-e^{-\beta\omega})$ factor provides a countervailing large inverse Boltzmann factor, leading the energy gain processes to make an appreciable contribution to $F^\mathrm{RIXS}_Q$. Because it is not possible to measure the exponentially suppressed response, we use a model computation to determine this $\omega <0$ contribution to $F^\mathrm{RIXS}_Q$.  The need to do this can be avoided in certain circumstances.  When we have polarization resolution in the outgoing photon (which we ourselves do not have here) {\it and} we work in the ultrashort core hole lifetime (UCL) approximation, we can use the relation \cite{shen2025}
\begin{equation}
    \chi''_{\hat{A}_R\hat{A}^\dagger_R}(\omega;k_o,\boldsymbol\epsilon_o;k_i,\boldsymbol\epsilon_i)=\chi''_{\hat{A}^\dagger_R\hat{A}_R}(\omega;k_i,\boldsymbol\epsilon_i;\boldsymbol\epsilon_o,k_o).
\end{equation}
This implies that under the UCL approximation, the RIXS gain intensity associated with a particular incoming and outgoing photons' momenta and polarization is related to the RIXS loss intensity arising from reversing the incoming and outgoing photons.

 We have thus solved the question of non-Hermiticity of the RIXS operator in detecting entanglement.  We now turn to the issue of the non-locality of the RIXS operator.  While the RIXS operators appear to be the sum over dipole operators defined on the individual sites, the presence of the energy denominator $1/(\omega_{in}-E_n+E_g+i\Gamma_c)$ is not a local operator per se -- unless one applies the ultra-short core hole lifetime approximation where information about the core hole dynamics encoded in $E_n$ is neglected.  While we will not use this approximation here we consider it in further detail in Sec.~IV A of \cite{supp}. However the locality of the RIXS operator is not necessary to use it to detect entanglement.  Losing locality simply means a simple relation such as Eqn.~\ref{eq:nQFI} no longer holds.
However one can still compute QFI bounds directly.  To determine the bound for wavefunctions with $m$-partite entanglement, one computes
\begin{equation}\label{QFIbound}
F^\mathrm{RIXS,max}_{Q,m-\text{partite}} = {\rm max}_{|\psi_{m-\text{partite}}\rangle}F_Q(|\psi\rangle\langle \psi|,\hat A) .
\end{equation}
Beyond making no assumptions on the locality of the RIXS operator, this approach also has the advantage of allowing one to account for symmetries in the system.  The formulation of the bound in Eqn.~\ref{eq:nQFI} makes no assumptions on any symmetry of the material (for example electron number conservation, (twisted) parity, or spin symmetry).  If one computes instead the bound as in Eqn.~\ref{QFIbound} one can simply account for symmetries by restricting the wavefunctions in the maximization of $F_Q$ to wavefunctions with the desired symmetry.  This in general makes for more sensitive detection of entanglement.  It does, however, come with added numerical cost (one has to compute directly the bounds in Eqn.~\ref{QFIbound}).  For our case of a 2-site dimer system, this is a manageable task.  Computing these bounds for larger systems is still doable: in Ref.~\cite{PhysRevLett.133.260202}, bounds were computed for wavefunctions with up to 8-partite entanglement over thermodynamically large systems.

In the following section, we show how to derive $F^\mathrm{RIXS}_Q$ and its corresponding bounds in the case of dimer iridates. This material features particularly well defined dimers with strong intra-dimer interactions and weak inter-dimer interactions \cite{Revelli2019resonant}. It therefore represents a particularly clear and simple case for illustrating QFI metrology in RIXS.

\begin{figure}[htp!]
	\centering
	\includegraphics[width=\linewidth]{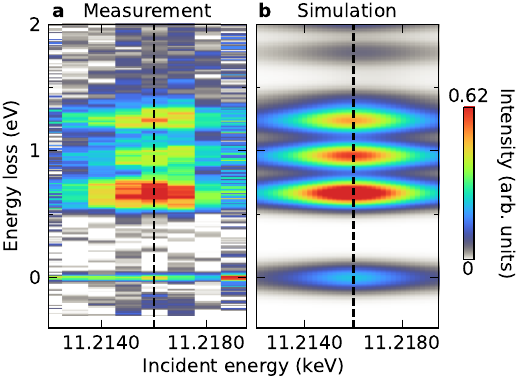}
	\caption{\textbf{Comparison of the incident energy dependence of the RIXS spectra between measurement and simulation.} \textbf{a} Measured RIXS spectra with varying incident energy at fixed momentum transfer $Q=(-0.5,0,18.94)$ in reciprocal lattice units (r.l.u.). The signals shown correspond to intra-$t_{2g}$ transitions. \textbf{b} Calculated incident-energy-dependent RIXS spectra at the same fixed momentum transfer. On both panels, the dashed line marks the incident energy used for the momentum-dependent scan in Fig. \ref{fig:CeLmap}.}
	\label{fig:CeEmap}
\end{figure}

\begin{figure}[htp!]
	\centering
	\includegraphics[width=\linewidth]{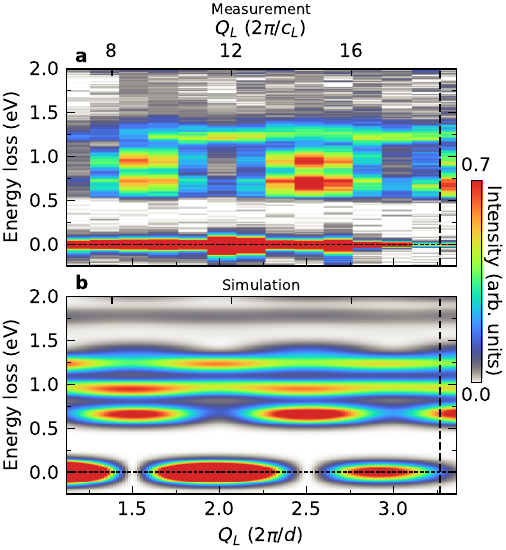}
	\caption{\textbf{Comparison of the $L$ dependence of the RIXS spectra between measurement and simulation. a} Measured $L$-dependent RIXS spectra showing periodic modulation, where the momentum transfer is $Q=(-0.5,0,L)$ r.l.u., and the incident energy $\omega_{\text{in}}$ is fixed to 11.216~keV. Only the intra-$t_{2g}$ transitions are presented since they are the dominating signals at this particular incident energy. \textbf{b} Calculated $L$-dependent RIXS spectra at the same fixed incident energy. $Q_L$ is the momentum transfer along the $L$ direction. For convenience, the same momentum scale is displayed in two units. On top, we use units of $2\pi/c_L$ where $c_L$ is the unit cell lattice constant along $L$ direction. On bottom, we use units of $2\pi/d$, where $d$ is the distance between the atoms that make up the dimer along the $L$ direction. On both panels, the dashed line marks the fixed momentum used for the incident photon energy-dependent scan in Fig. \ref{fig:CeEmap}.}
	\label{fig:CeLmap}
\end{figure}

\section*{Case study in iridate dimer materials}
As shown in Fig.~\ref{fig:overview}, Ba$_3$CeIr$_2$O$_9$ features two Ir sites in its basic structural motif hosting ten $5d$ electrons between them. It has been studied previously and has been shown to exhibit a clear Ir $L$-edge RIXS orbital signal, including a strong interference effect \cite{Revelli2019resonant}. Here we show that the processes that create a core hole at either one of these sites can be used to formulate a RIXS witness for the simplest form of multiparticle entanglement --- that of bipartite entanglement --- between the sites. 

\begin{figure}[htp!]
	\centering
	\includegraphics[width=\linewidth]{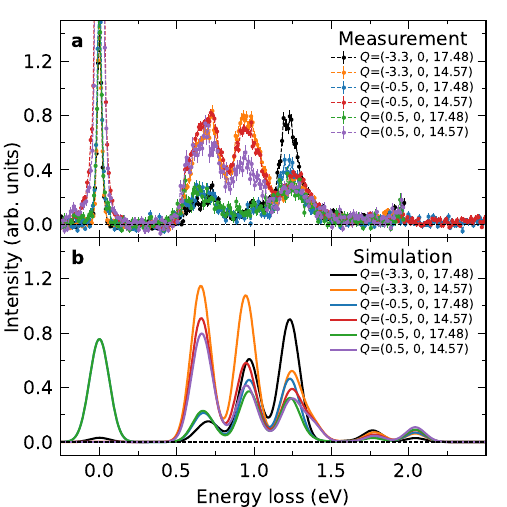}
	\caption{\textbf{Comparison of RIXS spectra between measurement and simulation at representative momentum transfers. a} Representative RIXS spectra at the indicated momentum transfers in r.l.u., and the incident energy $\omega_{\text{in}}$ is fixed to 11.216~keV. Only the intra-$t_{2g}$ transitions are presented since they are the dominant signal at this particular incident energy.  \textbf{b} Calculated RIXS spectra at the same momentum transfers and the same fixed incident energy. The model faithfully reproduces the trends seen in the experiment. The first two inelastic features around 0.7 and  0.9~eV are strong at $L=14.57$ and weak at $L=17.48$ due to constructive or destructive inter-site interference, respectively. The feature around 1.2~eV has different symmetry and the opposite trend in $L$.}
	\label{fig:CeOmap}
\end{figure}

\begin{figure*}[htp!]
	\centering
	\includegraphics[width=\textwidth]{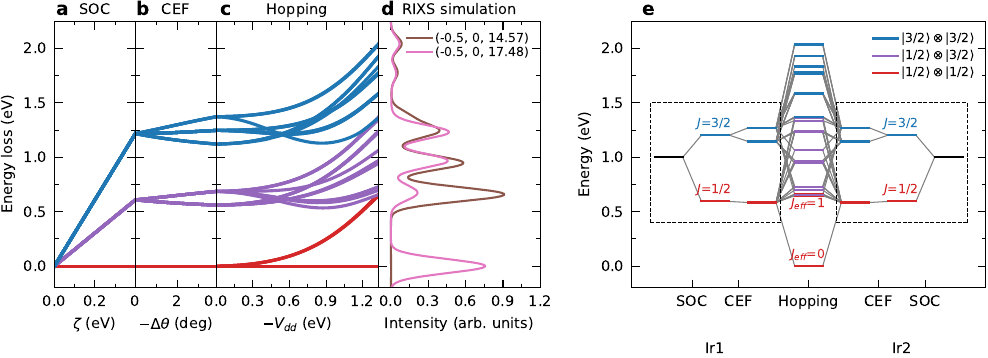}
	\caption{\textbf{Tracing the evolution of the eigenenergies from \gls*{ED} calculations. a} For a single Ir$^{4+}$ ion with octahedral coordination, the degenerate $t_{2g}$ orbital manifold is split into $J=1/2$ and $J=3/2$ multiplets due to the strong \acrfull*{SOC}. For two Ir$^{4+}$ ions, these make three combinations of $\ket{1/2}\otimes \ket{1/2}$ (red),  $\ket{1/2}\otimes \ket{3/2}$ (purple), and $\ket{3/2}\otimes \ket{3/2}$ (blue). \textbf{b} The inclusion of trigonal distortion induced \acrfull*{CEF} leads to the splitting of the $J=3/2$ state. \textbf{c)} The inter-atomic hopping mixes and rearranges all the states which contribute to the RIXS spectra in \textbf{d)}. \textbf{e)} The orbital-energy diagram summarizing \textbf{a}--\textbf{c} indicates that the ground state of the system is a $J_{\textrm{eff}}=0$ singlet state originating from the two interacting $J=1/2$ Ir doublets and the strongest RIXS peak at around 0.7~eV corresponds to the $J_{\textrm{eff}}=1$ triplet state.}
	\label{fig:level}
\end{figure*}

To apply our methodology to detect entanglement, we need to model the material-specific RIXS operator. This is done using the numerical toolkit EDRIXS \cite{WANG2019151} in combination with Hamiltonian parameter inference \cite{tnqm-ttj3} to provide the model parameters (including on-site Slater parameters, spin-orbit couplings and three hopping integrals $V_{dd\sigma}, V_{dd\pi}, V_{dd\delta}$, see Table~\ref{table:allparams}, Section V in the SM~\cite{supp}, Refs.~\onlinecite{Revelli2019resonant,Kugel2015Spin}) together with a single overall scale factor to reproduce the experimentally measured RIXS intensity. In fitting these parameters we restrict our focus on the energy range (0.2~eV, 2~eV) comprising the t$_{2g}$-manifold of states.  We exclude from consideration energies below 0.2~eV to avoid contamination from the elastic peak whose contribution to the QFI is properly zero --- see Eqn.~\ref{RIXS_QFI}.  This prevents instrumental broadening artifacts from contaminating our QFI estimates.

We first demonstrate that our modeling efforts are able to reproduce the measured RIXS intensity for this dimer material.  The main results are shown in Figs.~\ref{fig:CeEmap}, \ref{fig:CeLmap}, and \ref{fig:CeOmap}. It can be seen from the figures that the constructed RIXS operator indeed reproduces the essential features of the measured RIXS intensity, especially the major RIXS peaks below the energy loss gap around 2~eV, and these RIXS peaks are the dominant signal for incident energy around 11.216~keV.

To better understand the electronic characteristics of these major RIXS peaks at low energy losses, we trace the evolution of the eigenenergies with the \gls*{SOC}, the \gls*{CEF} due to trigonal distortion, and the hopping amplitude. This is shown in Fig.~\ref{fig:level}, where we can see that the ground state lies in the $t_{2g}$ orbital manifold, and the excitations below 2~eV are primarily intra-$t_{2g}$. This is a similar assignment to that made previously based on a model including only $t_{2g}$ orbitals \cite{Revelli2019resonant}.

\begin{figure}
    \centering
    \includegraphics[width=\linewidth]{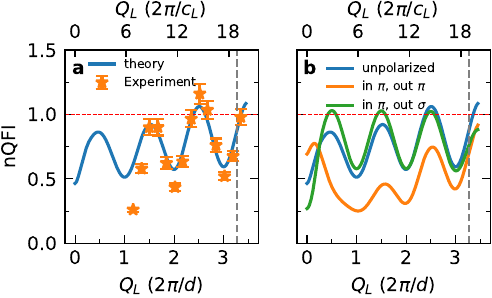}
    \caption{\textbf{The entanglement witness --- momentum dependence}. \textbf{a} comparison of simulation and measurement, for different momentum transfers $Q_L=(-0.5,0,L)$ r.l.u., at fixed incident energy $\omega_{\text{in}}=11.216$ keV. \textbf{b} Same setup as \textbf{a} but with final state polarization discrimination. The legend in the right panel documents the polarizations for the final state, where the $\pi$ polarization is in the incident plane, $\sigma$ polarization is out of the incident plane. The red dashed line on both panels is the nQFI threshold for 2-partite entanglement. For better comparison, the fixed momentum used for the energy-dependent scan on Fig. \ref{fig:CeQFI} is indicated with a dashed gray line.}\label{fig:CeExp}
\end{figure}

 With the RIXS operator at hand, we first calculate the QFI, $F^\mathrm{RIXS}_Q$ from Eq.~\ref{RIXS_QFI} at different incident photon energies and different momentum transfers.  We will see that not all energies and momentum detect entanglement. In order to incorporate the contribution at negative frequencies $\omega<0$, we use the model verified on the energy loss $\omega>0$ data to compute $I_{\text{RIXS}}(\omega)$. This is necessary because such processes are thermally suppressed so that they cannot be seen directly but the weighting of integral in Eq.~\eqref{RIXS_QFI} by $e^{-\beta\omega}$ leads them to contribute appreciably to the entanglement estimate.

In order to turn $F^\mathrm{RIXS}_Q$ into a predictor of entanglement we need to compute $F^\mathrm{RIXS}_{Q,\text{max},1}$, i.e. the maximal value $F^\mathrm{RIXS}_Q $ can have if the wavefunction of the system has no (i.e., 1-partite) entanglement between the two dimer sites.  To compute this maximum we restrict ourselves to wavefunctions that both respect twisted parity (the equivalence under exchange of the site indices when combined with a 180$^o$ rotation) and electron number conservation.  In the case of Ba$_3$CeIr$_2$O$_9$, there are 10 electrons in total shared among the two sites in the dimer.  We are restricting ourselves to the $t_{2g}$ manifold so there are six possible states $\{|B_u\rangle\}^6_{j=1}$ on each site. A generic wavefunction on either of the two sites is given by $\sum^6_j a_j |B\rangle_j$.  A 1-partite wavefunction for the dimer that respects parity must then take the form $|\psi_\alpha\rangle=|\alpha\rangle\otimes|\alpha\rangle$.  The maximal QFI of such wavefunctions, $F^{\mathrm{RIXS},\text{max}}_{Q,1-\text{partite}}$ is then computed by maximizing 
\begin{eqnarray}
F_Q(\alpha,\hat{A})&=&4\left(\bra{\psi_\alpha}\hat{A}_{R,\text{Re}}^2\ket{\psi_\alpha}-\bra{\psi_\alpha}\hat{A}_{R,\text{Re}}\ket{\psi_\alpha}^2\right) \quad\quad\quad \cr\cr
&& \llap{+}  4\left(\bra{\psi_\alpha}\hat{A}_{R,\text{Im}}^2\ket{\psi_\alpha}-\bra{\psi_\alpha}\hat{A}_{R,\text{Im}}\ket{\psi_\alpha}^2\right).
\end{eqnarray}
With this maximal 1-partite entanglement in hand, we define the normalized nQFI, $f^\text{RIXS}_Q$ as
\begin{equation}
    f^\mathrm{RIXS}_Q = \frac{F^\mathrm{RIXS}_Q}{F^\mathrm{RIXS,max}_{Q,1-\text{partite}}}
\end{equation}
Whenever $f^\mathrm{RIXS}_Q>1$, we can certify that the dimers are entangled.  We consider in more detail the construction of this bound in Secs.~IV B and C of \cite{supp}.

\begin{figure}[htp!]
	\centering
	\includegraphics[width=\linewidth]{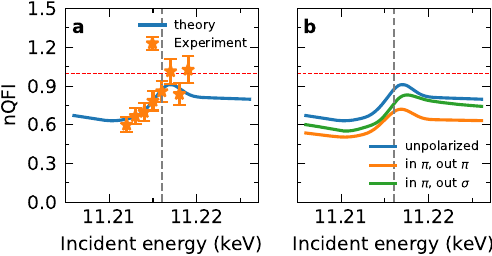}
	\caption{ {\bf The entanglement witness --- energy dependence.}  \textbf{a} comparison of simulation and measurement, for different incident photon energies at fixed momentum transfer $Q=(-0.5,0,18.94)$ r.l.u. \textbf{b} Same setup as \textbf{a} but with final state polarization discrimination. The legend in the right panel documents the polarizations for the final state, where the $\pi$ polarization is in the scattering plane, $\sigma$ polarization is out of the scattering plane. The red dashed line on both panels is the nQFI threshold for 2-partite entanglement.   For better comparison, the fixed energy used for the momentum-dependent scan on Fig. \ref{fig:CeExp} is indicated with a dashed gray line.
    }
	\label{fig:CeQFI}
\end{figure}

Firstly, we perform this calculation with the same incident energies and momentum transfers as the existing experimental data, as plotted in Fig.~\ref{fig:CeExp}. We can see that there are significant dependencies on momenta and incident photon energy.  The nQFI as a function of momentum has an oscillatory dependence that matches the oscillations in intensity in the t$_{2g}$ signal seen in Fig.~\ref{fig:CeLmap}b.  While it might seem obvious that the nQFI is largest when the RIXS intensity is largest, this is not a foregone conclusion as the nQFI is a normalized quantity and its actual value depends on a delicate balance between the RIXS intensity and the maximal intensity associated with 1-partite wavefunctions.  We also see in Fig.~\ref{fig:CeQFI}a that there appears to an optimal photon energy, 11.2157~keV, for which to detect entanglement.  Again this corresponds to a maximal intensity in the t$_{2g}$ RIXS response.
For reference, we provide both the unnormalized QFI as well as $F^{\text{RIXS,max}}_{Q,{\rm 1-partite}}$ in Sec.~IV C of \cite{supp}.

In addition to the results for Ba$_3$CeIr$_2$O$_9$, we also performed equivalent analysis for related iridium-dimer material Ba$_3$TaIr$_2$O$_9$, which host 11 electrons split between he Ir $5d$ orbitals. The results are presented in the Supplementary Information and are consistent with those for Ba$_3$CeIr$_2$O$_9$ \cite{supp}, supporting the generality of the applicability of our protocol.

Our scattering data, as is typical for hard x-ray RIXS, does not have out-going polarization discrimination.  However because polarization resolved RIXS is now becoming possible \cite{Gao2016toroidal, Kim2016collimating}, we repeat our analysis where incoming and outgoing polarization is fixed.  In Fig.~\ref{fig:CeExp}b we see how resolving polarization affects the values of the nQFI as a function of incident momentum.  Notably the RIXS intensity leads to a much larger nQFI when the polarization of the incoming photon is perpendicular to the outgoing photon, detecting entanglement for a larger set of momenta than either the unpolarized intensity or the case where the polarizations of the incoming and outgoing photon are different.

We similarly consider the polarization dependence of the nQFI as function of incident photon energy at a fixed momentum in Fig.~\ref{fig:CeQFI}b.  Here we see the nQFI of the unpolarized intensity is superior to either of polarization resolved nQFIs in detecting entanglement.  In fact we see that the two polarization resolved nQFIs never exceed the threshold of 1 needed to detect bipartite entanglement for any of the considered photon energies.

\section*{Conclusion and Discussion}

In this letter, we have generalized the protocol of using the QFI for entanglement detection to the case with a non-Hermitian operator, and applied it to iridate dimer systems using RIXS as the entanglement witness. 
Using careful parameter estimation for modeling the dimer material's Hamiltonian combined with a correct accounting for natural symmetries in the wavefunction, both the equivalence of the dimer sites together with electron number conservation, we are able to show the QFI indicates bipartite entanglement of the dimer sites.  We have also explored the role that momentum, incident energy and polarization play in detecting entanglement.  We have shown that entanglement detection is favored at certain momenta and energy, corresponding to maxima in the RIXS intensity.  We have also considered the case of polarization discriminated RIXS. Here we find a mixed result.  Polarization discrimination can help detect entanglement but is not a panacea.

Our result constitutes the first approach to experimentally detect orbital electronic quantum entanglement in a real quantum material. Experimental detection of quantum entanglement has been previously reported only for synthetic few-body systems \cite{Blatt_2005,Oberthaler_2008,GUHNE20091,Oberthaler_2010,PhysRevLett.109.020505,Cramer_2013,Jurcevic_2014,PhysRevLett.115.035302,Pitsios_2017,PhysRevA.98.052334,doi:10.1126/science.aau4963}, and for bosonic degrees of freedom in low-dimensional quantum magnets \cite{Ghosh_2003,PhysRevB.73.134404,PhysRevA.73.012110,PhysRevLett.99.087204,doi:10.1073/pnas.0703293104,PhysRevB.75.054422,Das_2013,Singh_2013,Sahling_2015,PhysRevResearch.2.043329,Tennant_PhysRevB.103.224434,PhysRevLett.127.037201, Scheie2024proximate}. Besides, quantum entanglement detection has been simulated theoretically for bosonic degrees of freedom in several 1D strongly correlated model systems out of equilibrium \cite{PhysRevLett.130.106902,neutron_from_trRIXS}. We have generalized the protocol of using the QFI for entanglement detection to non-Hermitian operators, enabling the direct detection of electronic quantum entanglement. This has enabled us to use RIXS spectra to infer the presence of entanglement, but also shows how other measured correlation functions of quantum materials that involve a non-Hermitian operator can be used to detect the presence of multipartite entanglement. This will be important for material classes such as the iridates considered here where inelastic neutron scattering is impractical.

\section*{Methods}
\label{sec:methods}

\textit{Sample synthesis}:
Single crystals of Ba$_3$CeIr$_2$O$_9$ were grown using the melt-solution technique in BaCl$_2$ flux. Stoichiometric quantities of BaCO$_3$, CeO$_2$, and IrO$_2$ were used as starting materials. After growth, the crystals were mechanically separated from the flux and washed with water to remove flux residues. The crystal possesses a $P6_{3}/mmc$ (No.~194) space group with lattice constants of $a = b = 5.9035(9)$~\AA{}, $c = 14.715(3)$~\AA{}, $\alpha=\beta=90^\circ$ and $\gamma=120^\circ$ \cite{Revelli2019resonant}. Within this crystal structure, the Ir atoms forming the dimer are separated by $d = 2.5361(7)$~\AA{} along the sample $c$-axis direction.

\textit{RIXS measurements}:
The RIXS experiments were performed at the 27-ID-B endstation of the Advanced Photon Source at Argonne National Laboratory. We used a spherically bent Si (844) diced analyzer and a Si (844) channel cut monochromator. The overall energy resolution was around 42~meV. The sample was mounted with its $H$ and $L$ reciprocal lattice vectors in the horizontal scattering plane and horizontally ($\pi$)-polarized incident x-rays were used. Data were collected at 9~K for Ba$_3$CeIr$_2$O$_9$ and at 11~K for Ba$_3$TaIr$_2$O$_9$. For all the \gls*{RIXS} data presented in the manuscript, a constant background has been subtracted using the energy-gain side followed by an angle-dependent self-absorption correction for each spectrum. The self-absorption correction was performed using methods described in Ref.~\cite{Miao2017high} following
\begin{equation}
    \frac{I_\textrm{RIXS}}{I_\textrm{raw}} = \frac{\sin \theta_o}{\sin \theta_i + \sin \theta_o }
\end{equation}
where $I_\textrm{RIXS}$ and $I_\text{raw}$ are scattering intensities with and without self-absorption and $\theta_i$ and $\theta_o$ are the angles of incidence and exit w.r.t.\ the sample surface. Although a more sophisticated approach would consider the photon polarization, that effect is small in this material, and can be neglected.

\textit{Exact diagonalization calculations}: 
The \gls*{ED} calculations for both iridates were performed using the EDRIXS software based on a two-Ir-site cluster \cite{WANG2019151}. For the Hamiltonian, we explicitly include the onsite Coulomb interactions and spin orbit coupling. In the fitting procedure, the direct hoppings between the neighboring Ir atoms are described by three Slater-Koster parameters, $V_{dd\sigma}$, $V_{dd\pi}$ and $V_{dd\delta}$, and together with the Slater parameters $F^2_{dd}$, $F^4_{dd}$, $F^2_{dp}$, $G^1_{dp}$, $G^3_{dp}$ and the spin-orbit couplings $\zeta$ and $\zeta_c$ are treated as fit parameters; in the subsequent greedy refinement step we additionally allow $10D_q$ and the core-hole broadening $\Gamma_c$ to vary. The trigonal distortion is included through the approach described in Ref.~\onlinecite{Kugel2015Spin}, in which $\theta$, the Ir-Ir-O angle, is fixed to the experimental value of the material, while $\kappa$ is kept fixed throughout the fitting. Further details of the Hamiltonian are discussed in Supplementary Section III \cite{supp}. Regarding the Hartree-Fock basis, we consider $t_{2g}^{10}e_g^{0}$ (and for Ba$_3$TaIr$_2$O$_9$, also $t_{2g}^{9}e_g^{1}$) configurations only, focusing on the low energy loss sector with negligible $e_g$ occupancy. Extending to at most one $e_g$ electron in the model demonstrates that the octahedral crystal field splitting can be seen directly in the different resonant energies of the $t_{2g}$ and $e_g$ manifolds. For Ba$_3$CeIr$_2$O$_9$ we further restrict to the $e_g^{0}$ sector. The richly-structured experimental data allow us to find a combination of parameters that can reproduce the collected \gls*{RIXS} data, the full list of which is presented in Table~\ref{table:allparams}. We use a Bayesian optimization-based inverse scattering method following Ref.~\onlinecite{Lajer2025Ham} to obtain the model parameters through a two-step procedure consisting of Gaussian process regression followed by greedy refinement. We outline our approach in Supplementary Section ~V \cite{supp}. The \gls*{RIXS} spectra are calculated using the Kramers-Heisenberg formula in the dipole approximation with the experimental geometry explicitly considered. The calculations overall show good agreement with experiment.

Our model follows the general framework of Ref.~\onlinecite{Revelli2019resonant}, but we implement it in a form that can, when needed, include finite $e_g$ occupancy in addition to $t_{2g}$ occupancy. Allowing $e_g$ configurations improves agreement with experiment over a wider range of energy loss and is essential for systems where the electron count makes a strictly $t_{2g}$-restricted description too limiting, such as the Ta-based dimer discussed in Supplementary Section VI (11 electrons distributed over the two Ir $d$ shells). For Ba$_3$CeIr$_2$O$_9$, however, we restrict the subsequent QFI analysis to the $t_{2g}$ sector to retain computational tractability, while still using a parameterization appropriate to our present Hamiltonian. Spin-orbit coupling can admix $t_{2g}$ and $e_g$ character at the $\sim 20\%$ level \cite{Stamokostas2018Mixing}, which motivates the more general formulation.

\section*{Data availability}
The RIXS data generated in this study have been deposited in the Zenodo database under accession code [to be assigned].

\section*{Code availability}
The exact diagonalization is done using the codebase EDRIXS \cite{WANG2019151}. The code used in this study is available from the authors upon reasonable request.

\begin{table*}
\caption{Full list of parameters used in the \gls*{ED} calculations for Ba$_3$CeIr$_2$O$_9$. The \gls*{CEF} parameters such as $\theta$ and $\kappa$ are defined in Ref.~\onlinecite{Kugel2015Spin}. Here, $\zeta$ is the spin orbit coupling parameter for the Ir $5d$ orbitals while the one for the $2p$ core orbitals is 
fitted to be 1142.86~eV (the theoretical atomic model value is 1140.332~eV). The inverse core-hole lifetime is inferred to be 2.53~eV (the atomic value being 2.47~eV) and the final-state energy loss spectra are broadened using a Gaussian function with a full-width at half-maximum of 0.15~eV. All parameters except $\theta$ and $\kappa$ are in units of eV.}
\begin{ruledtabular}
\begin{tabular}{cccccccc}
\multicolumn{4}{c}{Crystal field splitting and spin orbit coupling} & & \multicolumn{3}{c}{Hopping integrals}\\
10$D_q$ & $\theta$ & $\kappa$ & $\zeta$ & & $V_{dd\sigma}$ & $V_{dd\pi}$ & $V_{dd\delta}$\\
3.23 & 50.86 & 0.1 & 0.41 & & -1.34 & -0.004 & 0.38\\
\hline
\multicolumn{4}{c}{Core-hole potential} & & \multicolumn{3}{c}{On-site Coulomb interactions}\\
$F^0_{dp}$ & $F^2_{dp}$ & $G^1_{dp}$ & $G^3_{dp}$ & & $F^0_{dd}$ & $F^2_{dd}$ & $F^4_{dd}$\\ 
2.584 & 0.266 & 1.199 & 0.104 & & 2.736 & 2.705 & 0.529\\
\end{tabular}
\end{ruledtabular}
\label{table:allparams}
\end{table*}

\section*{References}
\bibliography{main}

\section*{Acknowledgments}
Work at Brookhaven and Harvard was supported by the U.S.\ Department of Energy, Office of Science, Office of Basic Energy Sciences, under Award Number DE-SC0012704. This research used resources of the Advanced Photon Source, a U.S.\ Department of Energy (DOE) Office of Science User Facility at Argonne National Laboratory and is based on research supported by the U.S.\ DOE Office of Science-Basic Energy Sciences, under Contract No. DE-AC02-06CH11357.

\section*{Author contributions}
T.R., Y.S., M.L.\& R.M.K.\ performed the calculations. Y.S., S.F.R.T., J.S., W.H., M.H.U., D.C., M.M., and M.P.M.D.\ performed the measurements. P.B.\ prepared the crystals. The paper was written by T.R., Y.S., M.L., M.M., M.P.M.D. \& R.M.K. with input from all co-authors.

\section*{Additional Information}
Correspondence and requests for materials should be addressed to Y.S., M.L., M.P.M.D, or R.M.K.

\section*{Competing financial interests}
The authors declare no competing interests.

\end{document}


\title{Witnessing Quantum Entanglement Using Resonant Inelastic X-ray Scattering\\Supplementary Information}

\author{Tianhao Ren}
\affiliation{Condensed Matter Physics and Materials Science Division, Brookhaven National Laboratory, Upton, New York 11973, USA}

\author{Yao Shen}
\email{yshen@iphy.ac.cn}
\affiliation{Condensed Matter Physics and Materials Science Division, Brookhaven National Laboratory, Upton, New York 11973, USA}
\affiliation{Beijing National Laboratory for Condensed Matter Physics,
Institute of Physics, Chinese Academy of Sciences, Beijing 100190, China}

\author{Marton Lajer}
\email{mlajer@bnl.gov}
\affiliation{Condensed Matter Physics and Materials Science Division, Brookhaven National Laboratory, Upton, New York 11973, USA}

\author{Sophia F. R. TenHuisen}
\affiliation{Department of Physics, Harvard University, Cambridge, Massachusetts 02138, USA}

\author{Jennifer Sears}
\author{Wei He}
\affiliation{Condensed Matter Physics and Materials Science Division, Brookhaven National Laboratory, Upton, New York 11973, USA}

\author{Mary H. Upton}
\author{Diego Casa}
\affiliation{Advanced Photon Source, Argonne National Laboratory, Argonne, Illinois 60439, USA}

\author{Petra Becker}
\affiliation{Section Crystallography, Institute of Geology and Mineralogy, University of Cologne, 50939 K\"{o}ln, Germany}

\author{Matteo Mitrano}
\affiliation{Department of Physics, Harvard University, Cambridge, Massachusetts 02138, USA}

\author{Mark P. M. Dean}
\email{mdean@bnl.gov}
\affiliation{Condensed Matter Physics and Materials Science Division, Brookhaven National Laboratory, Upton, New York 11973, USA}
\author{Robert M. Konik}
\email{rmk@bnl.gov}
\affiliation{Condensed Matter Physics and Materials Science Division, Brookhaven National Laboratory, Upton, New York 11973, USA}

\date{\today}

\maketitle

\newcommand\tlc[1]{\texorpdfstring{\lowercase{#1}}{#1}}

\section{Dipole Transition Operators}

In the formula for the RIXS intensity presented in the main text, we encounter the transition operators $\mathcal{D}_i(\mathbf{k}_i,\bm{\epsilon}_i)$, $\mathcal{D}_f(\mathbf{k}_o,\bm{\epsilon}_o)$. Under the dipole approximation, these can be written as \cite{WANG2019151},
\begin{equation}
    \begin{split}
        & \hat{\mathcal{D}}_i=\sum_{a=x,y,z} \epsilon_{i,a} \hat{T}_{i,a}, \quad \hat{\mathcal{D}}_o^{\dagger}=\sum_{a=x,y,z} \epsilon_{o,a}^{*} \hat{T}_{o,a}^{\dagger}, \\
        & \hat{T}_{i,x}=e^{\mathrm{i} \mathbf{k}_i \cdot \mathbf{R} } \hat{x}_R , \quad \hat{T}_{i,y}=e^{\mathrm{i} \mathbf{k}_i \cdot \mathbf{R} } \hat{y}_R , \quad \hat{T}_{i,z}=e^{i \mathbf{k}_i \cdot \mathbf{R} } \hat{z}_R , \\
        & \hat{T}_{o,x}=e^{i \mathbf{k}_o \cdot \mathbf{R} } \hat{x}_R , \quad \hat{T}_{o,y}=e^{i \mathbf{k}_o \cdot \mathbf{R} } \hat{y}_R , \quad \hat{T}_{o,z}=e^{i \mathbf{k}_o \cdot \mathbf{R} } \hat{z}_R ,
    \end{split}
\end{equation}
where $\bm{R}$ is the coordinate for scattering site-$R$, and $\hat{x}_R$, $\hat{y}_R$, $\hat{z}_R$ are position operators of electrons bound to the site-$R$.

\section{Quantum Fisher Information and Susceptibilities}
 We consider a Hermitian operator $\hat{A}$. For a pure state, the QFI is expressible as \cite{pezze2014quantum}
\begin{equation}
\label{eq:pureQFI}
    F_Q(\rho, \hat{A})=4\left(\langle\psi|\hat{A}^2 |\psi\rangle-\langle\psi|\hat{A}| \psi\rangle^2\right).
\end{equation}
If instead of a pure state, the quantum state is described by a thermal density matrix given by $\rho = \sum_k\lambda_k|k\rangle\langle k|$, where $\lambda_k$ is the $k$-dependent Boltzmann factor, the QFI is expressible as \cite{pezze2014quantum}
\begin{equation}
    F_Q(\rho, \hat{A}) = 2\sum_{k,k'}\frac{(\lambda_k-\lambda_{k'})^2}{\lambda_k+\lambda_{k'}}|\langle k|\hat{A}|k'\rangle|^2.
\end{equation}
~\\
To relate the QFI for a thermal ensemble to the dynamic susceptibility with a Hermitian operator $\hat{A}$ \cite{hauke2016}, we express the dynamic susceptibility $\chi''_{AA}$ in terms of the Lehmann representation:
\begin{equation}
    \chi''_{AA}(\omega)=\frac{1}{Z} \sum_{m, n} e^{-\beta E_m}\left(1-e^{-\beta \omega}\right)\left|\langle m|\hat{A}| n\rangle\right|^2 \delta\left(\omega+E_m-E_n\right).
\end{equation}
Using the fact that
\begin{equation}
    \int_{-\infty}^{\infty} \mathrm{d} \omega \tanh \left(\frac{\beta \omega}{2}\right) \delta\left(\omega+E_m-E_n\right)=\tanh \left(\frac{\beta\left(E_n-E_m\right)}{2}\right)=\frac{\lambda_m-\lambda_n}{\lambda_m+\lambda_n},
\end{equation}
we can easily verify that
\begin{equation}
    F_Q(\rho,\hat{A})=2\int_{-\infty}^{\infty} \mathrm{d} \omega \tanh \left(\frac{\beta \omega}{2}\right)\chi''_{AA}(\omega)=4\int_{0}^{\infty} \mathrm{d} \omega \tanh \left(\frac{\beta \omega}{2}\right)\chi''_{AA}(\omega),
\end{equation}
where the fact $\chi''_{AA}(\omega)=-\chi''_{AA}(-\omega)$ is used. We can generalize the above formulation to a non-Hermitian operator $\hat{A}$, in which case the sum of the QFIs of the real and imaginary parts of the operator $\hat A$ (both of which are Hermitian) is given by 
\begin{equation}
    \begin{split}
        F_Q(\rho,\hat{A}_{\text{Re}})+F_Q(\rho,\hat{A}_{\text{Im}})&=2\int_{-\infty}^{\infty}\mathrm{d}\omega\tanh\left(\frac{\beta\omega}{2}\right)\left[ \chi''_{A_{\text{Re}}A_{\text{Re}}}(\omega)+\chi''_{A_{\text{Im}}A_{\text{Im}}}(\omega)\right] \\
        &=\int_{-\infty}^{\infty}\mathrm{d}\omega\tanh\left(\frac{\beta\omega}{2}\right) \left[ \chi''_{AA^{\dagger}}(\omega)+\chi''_{A^{\dagger}A}(\omega)\right] \\
        &=2\int_{0}^{\infty}\mathrm{d}\omega\tanh\left(\frac{\beta\omega}{2}\right) \left[ \chi''_{AA^{\dagger}}(\omega)+\chi''_{A^{\dagger}A}(\omega)\right],
    \end{split}
\end{equation}
where we have used the fact that $\chi''_{AA^{\dagger}}(\omega)=-\chi''_{A^{\dagger}A}(-\omega)$. \\

\begin{figure}[htp!]
	\centering
	\includegraphics[width=0.8\linewidth]{unit_cell.png}
        \caption{\textbf{Unit cell of Ba$_3$CeIr$_2$O$_9$.} Ba is shown as gray, Ce as green, Ir as red, and O as blue. The red, green, and blue arrows, respectively, show the unit cell directions Thin black lines define the size of the unit cell. Octahedra making up the dimers, containing two Ir atoms separated by $d$, are outlined as red polyhedra.}
	\label{fig:unit_cell}
\end{figure}

\section{The form of the Hamiltonian for the iridate dimer compounds}
Schematically, the Hamiltonian describing the dimer is written as a sum of a quadratic and a quartic piece:
\begin{equation}
\hat{H}_\text{full} = \hat{H}_2 + \hat{H}_4.
\end{equation}
Below we consider each part in turn.
\subsection{Quadratic terms}

We construct a model following the derivation in Ref.~\cite{Kugel2015Spin}, who considered the crystal field and hopping terms relevant to face-sharing octahedra as shown in Fig.~\ref{fig:unit_cell}. Their starting ``global'' coordinate system puts the $z$ axis parallel to the threefold/trigonal axis through the centers of the face-sharing octahedra, which we label $s=1,2$. $x$ then lies in the plane perpendicular to $z$ maximizing its projection towards the nearest O atom. The degree of trigonal distortion is parameterized by $\theta$, which is defined as the angle between $z$ and the shared O atom \footnote{See Fig.~2(a) of Ref.~\cite{Kugel2015Spin}, which illustrates the coordinate system and definition of $\theta$.}. Following the standard EDRIXS $d$-shell orbital order of $\left|3z^{2}-r^{2}\right\rangle ,\left|xz\right\rangle ,\left|yz\right\rangle ,\left|x^{2}-y^{2}\right\rangle ,\left|xy\right\rangle$, the crystal field matrix is

\begin{equation}
C^{\left(s\right)}=10Dq\left(\begin{array}{ccccc}
-\frac{6}{5}a_{4}+\frac{2}{3}a_{2}&0&0&0&0\\
0 & \frac{4}{5}a_{4}+\frac{1}{3}a_{2} & 0 & (-1)^{s}\frac{b}{2} & 0\\
0 & 0 & \frac{4}{5}a_{4}+\frac{1}{3}a_{2} & 0 & (-1)^{s+1}\frac{b}{2}\\
0 & (-1)^{s}\frac{b}{2} & 0 & -\frac{1}{5}a_{4}-\frac{2}{3}a_{2} & 0\\
0 & 0 & (-1)^{s+1}\frac{b}{2} & 0 & -\frac{1}{5}a_{4}-\frac{2}{3}a_{2}
\end{array}\right),
\end{equation}
with $10Dq$ denoting the cubic crystal field splitting and where we have defined the parameters,
\begin{align}
a_{2} & =\frac{27}{35}\kappa\left(3\cos^{2}\theta-1\right)-a_{0}\nonumber; \\
a_{4} & =-\frac{3}{2}\left(\frac{5}{2}\cos^{4}\theta-\frac{15}{7}\cos^{2}\theta+\frac{3}{14}\right)\nonumber; \\
b & =3\sin^{3}\theta\cos\theta .
\end{align}

The block form of the crystal-field-hopping
(one-body/orbital) Hamiltonian is
\begin{equation}
H_{cf+V}^{\left(\text{glob}\right)}=\left(\begin{array}{cc}
C^{\left(1\right)} & V^{\left(1\rightarrow2\right)}\\
V^{\left(2\rightarrow1\right)} & C^{\left(2\right)}
\end{array}\right).
\end{equation}
where the $V$-submatrices take the form
\begin{align}
V^{\left(1\rightarrow2\right)} & =\mathrm{diag}\left(V_{dd}^{\sigma},V_{dd}^{\pi},V_{dd}^{\pi},V_{dd}^{\delta},V_{dd}^{\delta}\right)\nonumber ;\\
V^{\left(2\rightarrow1\right)} & =V^{\left(1\rightarrow2\right)},
\end{align}

We now perform a useful ``local'' basis transform with the primary axis along the trigonal symmetry axis such that there is no mixing between $x^2-y^2$ and  $3x^2-r^2$ states
\begin{equation}
H_{cf+V}^{\left(\text{loc}\right)}=\left(U^{\left(t2o\right)}\right)^{\dagger}H_{cf+V}^{\left(\text{glob}\right)}U^{\left(t2o\right)}.
\end{equation}
Here $U^{t2o}$ is defined by
\begin{equation}
U^{\left(t2o\right)}=\left(\begin{array}{cc}
U^{\left(1\right)} & 0\\
0 & U^{\left(2\right)}
\end{array}\right),
\end{equation}
with $U^{s=1,2}$ defined by
\begin{equation}
U^{\left(s\right)}=\left(\begin{array}{ccccc}
0 & \sqrt{\frac{1}{3}} & \sqrt{\frac{1}{3}} & 0 & \sqrt{\frac{1}{3}}\\
\left(-1\right)^{s}\sqrt{\frac{2}{3}} & \left(-1\right)^{s}\frac{\sqrt{2}}{6} & \left(-1\right)^{s}\frac{\sqrt{2}}{6} & 0 & -\left(-1\right)^{s}\frac{\sqrt{2}}{3}\\
0 & \left(-1\right)^{s}\sqrt{\frac{1}{6}} & -\left(-1\right)^{s}\sqrt{\frac{1}{6}} & \left(-1\right)^{s}\sqrt{\frac{2}{3}} & 0\\
\sqrt{\frac{1}{3}} & -\frac{1}{3} & -\frac{1}{3} & 0 & \frac{2}{3}\\
0 & \sqrt{\frac{1}{3}} & -\sqrt{\frac{1}{3}} & -\sqrt{\frac{1}{3}} &0
\end{array}\right).
\end{equation}

We now include spin by expanding the orbital space as $\mathcal{H}_\text{full}=\mathcal{H}_\text{orb}\otimes C_{2}$. In particular, $H_{\text{cf}+V}^{\left(\text{loc}\right)}$ gets extended to a $20\times20$
matrix $H_\text{\text{cf}+V,spin}^{\left(\text{loc}\right)}$. Spin-orbit
coupling is included after defining its matrix elements in the local frame. For both
sites, the matrix in the local (orbital and local spin) frame takes
the form
\begin{equation}
\zeta\mathbf{L}\cdot\mathbf{S}=\zeta\left(\mathcal{U}^\text{c2r}\right)^{\dagger}\left[L_{z}\otimes S_{z}+\frac{1}{2}\left(L_{+}\otimes S_{-}+L_{-}\otimes S_{+}\right)\right]\mathcal{U}^\text{c2r},
\end{equation}
where $\mathcal{U}^\text{c2r}$ transforms between the complex and real spherical harmonic orbital basis. Since the reference spin directions are aligned with the local coordinates, this SOC is not yet in the same basis as the other terms of the Hamiltonian.

To make the SOC consistent with the other terms, we further transform the SOC operator by a spin-dependent transformation, so that it is expressed in a ``global'' spin basis. To this end, we introduce
\begin{equation}
W^\text{spin}=\left(\begin{array}{cc}
\mathbb{1}_{5}\otimes d^{\left(1\right)} & \\
 & \mathbb{1}_{5}\otimes d^{\left(2\right)}
\end{array}\right),
\end{equation}
with $d^{\left(s\right)}\equiv d\left(\alpha^{\left(s\right)},\beta^{\left(s\right)},\gamma^{\left(s\right)}\right)$
and
\begin{equation}
d\left(\alpha,\beta,\gamma\right)=\left(\begin{array}{cc}
e^{-\frac{i}{2}\left(\alpha+\gamma\right)}\cos\frac{\beta}{2} & -e^{-\frac{i}{2}\left(\alpha-\gamma\right)}\sin\frac{\beta}{2}\\
e^{\frac{i}{2}\left(\alpha-\gamma\right)}\sin\frac{\beta}{2} & e^{\frac{i}{2}\left(\alpha+\gamma\right)}\cos\frac{\beta}{2}
\end{array}\right)\label{dtransfpdef},
\end{equation}
so that the quadratic Hamiltonian takes the form
\begin{equation}
H_2^\text{O-loc,S-glob}=H_\text{cf+V}^{\left(\text{loc}\right)}+\zeta\left(W^\text{spin}\right)^{\dagger}\mathbf{L}\cdot\mathbf{S}W^\text{spin}.\label{FinalQuadHam}
\end{equation}
We note that the Euler angles defining the spin rotation read as follows: $\alpha^{\left(1\right)}=45{^\circ}$,
$\beta^{\left(1\right)}=54.74{^\circ}$, $\gamma^{\left(1\right)}=0{^\circ}$ and
$\alpha^{\left(2\right)}=45{^\circ}$, $\beta^{\left(2\right)}=54.74{^\circ}$,
$\gamma^{\left(2\right)}=180{^\circ}$.

\subsection{Quartic interaction terms}

On-site and core-valence Coulomb interactions take the form
\begin{equation}
\hat{H}_4 = \sum_{\alpha,\beta,\gamma,\delta,\sigma,\sigma^\prime}
U_{\alpha\sigma,\beta\sigma^\prime,\gamma\sigma^\prime,\delta\sigma}
\hat{f}^{\dagger}_{\alpha\sigma}
\hat{f}^{\dagger}_{\beta\sigma^\prime}
\hat{f}_{\gamma\sigma^\prime}\hat{f}_{\delta\sigma},
\end{equation}
where $\alpha$, $\beta$, $\gamma$, and $\delta$ are orbital indices, $\sigma$ is the spin index, and $\hat{f}^{\dagger}$ ($\hat{f}$) are the creation (annihilation) operators. Following the standard practice in RIXS, we include on-site terms $F^0_{dd}$, $F^2_{dd}$, $F^4_{dd}$, and core-valence terms $F^0_{dp}$, $F^2_{dp}$, $G^1_{dp}$, and $G^3_{dp}$ using the standard Slater parameterization as provided in the EDRIXS package \cite{WANG2019151}.
In our approximation, the tensor $U$ has nonzero matrix elements between orbitals localized on the same single site.

\subsection{Implementation of Twisted Parity Symmetry}
In determining the QFI bounds for the barium iridate, we insist that the dimer wavefunctions respect a twisted parity symmetry, a $\mathbb{Z}_2$ symmetry of the face-sharing dimer that is a \emph{twisted} version of site-exchange parity.
The origin of the twist is that the two octahedra are related by site interchange together with a $C_2$ rotation about the trigonal axis, so that exchanging the two sites also exchanges their local spin frames.
In the quadratic Hamiltonian of Eq.~(\ref{FinalQuadHam}) we have adopted an \emph{orbital-local, spin-global} single-particle basis: the orbital degrees of freedom are rotated by $U^{(t2o)}$ to a convenient local orbital frame, while the spinors on each site are rotated by the site-dependent SU(2) matrix $d^{(s)}$ (collected into $W^{\rm spin}$) so that both sites share a common \emph{global} spin quantization convention.
In this basis the intersite hopping is diagonal in spin (i.e.\ spin independent), whereas spin-orbit coupling appears as the rotated operator
$\zeta (W^{\rm spin})^{\dagger}\,\mathbf{L}\!\cdot\!\mathbf{S}\,W^{\rm spin}$.

In the orbital-local, spin-global basis, the conserved parity operator that commutes with the Hamiltonian is not the plain site swap $1\leftrightarrow2$, but instead
\begin{equation}
\tilde{P}_{\rm glob}
=
\begin{pmatrix}
0 & -i\,\mathbb{1}_{5}\otimes \big(d^{(2)}\big)^{\dagger} d^{(1)}\\[2pt]
+i\,\mathbb{1}_{5}\otimes \big(d^{(1)}\big)^{\dagger} d^{(2)} & 0
\end{pmatrix},
\qquad
\tilde{P}_{\rm glob}^{\dagger}=\tilde{P}_{\rm glob},
\qquad
\tilde{P}_{\rm glob}^{2}=\mathbb{1}.
\label{eq:Ptwist_global}
\end{equation}
The relative spinor rotation $\big(d^{(2)}\big)^{\dagger} d^{(1)}$ is fixed by the Euler angles used in Eq.~(\ref{dtransfpdef}); in our geometry the two sites differ by $\Delta\gamma=\pi$, so that $\big(d^{(2)}\big)^{\dagger} d^{(1)}$ reduces to a $\pi$ rotation about $z$ in the spinor representation.
The additional factor of $\mp i$ in Eq.~(\ref{eq:Ptwist_global}) is \emph{not} an arbitrary decoration.  Rather it is required for $\tilde{P}_{\rm glob}$ to be a Hermitian involution and, in the chosen single-particle phase conventions, it is the unique choice (up to an overall sign) that yields $[\tilde{P}_{\rm glob},H_2^\text{O-loc,S-glob}]=0$.

It is instructive to compare to the \emph{orbital-local, spin-local} basis, in which the spin quantization axis is chosen separately on each site and aligned with the corresponding local coordinate frame.
In that representation the \emph{rotation} part of the twist disappears (because the local spin labels are transported with the sites), but the commuting parity operator still has the characteristic off-diagonal $\pm i$ structure,
\begin{equation}
\tilde{P}_{\rm loc}
=
\begin{pmatrix}
0 & -i\,\mathbb{1}_{5}\otimes \mathbb{1}_{2}\\[2pt]
+i\,\mathbb{1}_{5}\otimes \mathbb{1}_{2} & 0
\end{pmatrix}.
\label{eq:Ptwist_local}
\end{equation}

 In our setup, we transform the full one-particle Hamiltonian directly from the orbital-local, spin-global basis to a \emph{phase-corrected spin-local} basis using a block-diagonal spinor transformation,
\begin{equation}
T_{\rm spin} \equiv 
\mathrm{diag}\!\Big(\mathbb{1}_{5}\otimes t^{(1)},\ \mathbb{1}_{5}\otimes t^{(2)}\Big),
\qquad
t^{(1)} = d(0,0,0),\qquad
t^{(2)} = i\,d(0,0,\pi),
\label{eq:Tspin_def}
\end{equation}
and conjugate the quadratic Hamiltonian as
\begin{equation}
H_2^\text{O-loc,S-glob} \ \mapsto\ H_2 = T_{\rm spin}^{\dagger}\,H_2^\text{O-loc,S-glob}\,T_{\rm spin}.
\label{eq:H_work_transform}
\end{equation}
An analogous block-diagonal transformation is applied to the core subspace, so that the full valence plus core one-body Hamiltonian is consistently expressed in the same basis.

In the resulting working basis, the parity operator takes the conventional form
\begin{equation}
P
=
\begin{pmatrix}
0 & \mathbb{1}_{5}\otimes \mathbb{1}_{2}\\
\mathbb{1}_{5}\otimes \mathbb{1}_{2} & 0
\end{pmatrix},
\qquad
\big[H_2,P\big]=\big[\hat{H}_\text{full},\hat{P}\big]=0,
\label{eq:Pswap_work}
\end{equation}
so that the Hamiltonian can be block-diagonalized into even and odd parity sectors using standard parity eigenstates. This is the representation employed throughout the subsequent exact-diagonalization and variational calculations.

In the Fock space, this parity operator acts on one-partite states as
\begin{equation}
\hat{P}\ket{\psi_1}\otimes\ket{\psi_2}=(-1)^{N_1 N_2}\ket{\psi_2}\otimes\ket{\psi_1} ,
\end{equation}
where $N_1$ and $N_2$ are the electronic occupations of site 1 and site 2, respectively. Due to this extra fermionic sign, the symmetric states $\ket{\psi}\otimes\ket{\psi}$ with $N_1=N_2=5$ reside in the odd parity sector.

Note that in this spin-local representation, the intersite hopping becomes spin dependent: expressing both sites in their own local spin frames introduces the relative spinor rotation into the hopping blocks (schematically, $V\mapsto V\otimes \sigma_z$), so the hopping is no longer diagonal in spin.

\section{Determining the QFI Bounds for $m$-partite states}

In this section we consider how to form the QFI bounds.  We do so first demonstrating how one would proceed using the ultra short-core hole lifetime approximation.  We then turn to the approach used in this paper: symmetry enhanced bounds.  This method takes into account the symmetries of the system (the number of electrons on each dimer and the twisted parity symmetry of the two iridate sites making up each dimer) and allows one to arrive at more refined bounds.

\subsection{Determining the bounds under the ultra short-core hole lifetime approximation}

The standard way one determines the presence of entanglement from the measurement of a two point function of the Hermitian operator, $A=\sum^N_{i=1}\hat A_i$, is the inequality 
\begin{equation}
\label{eq:nQFI}
    \frac{F_Q(\rho,\hat{A})}{\sum_{i=1}^N(\Delta a_i)^2}>m, \quad \Delta a_i=a_{i,\text{max}}-a_{i,\text{min}},
\end{equation}
where $a_{i,\text{max/min}}$ are the maximum and minimum eigenvalues of the individual site operators of which $\hat A$ is a sum over and the QFI associated with the operator $\hat A$ is given in Eqn.~2 of the main text.

In our case we are not able to construct the QFI of a single Hermitian operator but instead are able to compute the sum of the QFIs associated with two different operators, namely the real, $\hat A_{R, \rm Re}$, and imaginary, $\hat A_{R, \rm Im}$ parts of the RIXS operator, $\hat A_{R}$, given in Eqn.~6 of the main text.  We are able to relate the sum of these two QFIs to a weighted integral over the measured RIXS intensity, $I_\text{RIXS}(\omega)$.

The ability to determine the presence of entanglement from the sum of these two QFIs is governed by the following set of inequalities:
\begin{equation}
    \text{min}\left\{ \frac{F_Q(\rho,\hat{A}_{\text{Re}})}{\sum_{i=1}^N(\Delta a_{i,\text{Re}})^2}, \frac{F_Q(\rho,\hat{A}_{\text{Im}})}{\sum_{i=1}^N(\Delta a_{i,\text{Im}})^2} \right\}\leqslant \frac{F_Q(\rho,\hat{A}_{\text{Re}})+F(\rho,\hat{A}_{\text{Im}})}{\sum_{i=1}^N\left[(\Delta a_{i,\text{Re}})^2+(\Delta a_{i,\text{Im}})^2\right]} \leqslant \text{max}\left\{ \frac{F_Q(\rho,\hat{A}_{\text{Re}})}{\sum_{i=1}^N(\Delta a_{i,\text{Re}})^2}, \frac{F_Q(\rho,\hat{A}_{\text{Im}})}{\sum_{i=1}^N(\Delta a_{i,\text{Im}})^2} \right\}.
\end{equation}
Therefore by this inequality if
\begin{equation}\label{QFIsum}
\frac{F_Q(\rho,\hat{A}_{\text{Re}})+F(\rho,\hat{A}_{\text{Im}})}{\sum_{i=1}^N\left[(\Delta a_{i,\text{Re}})^2+(\Delta a_{i,\text{Im}})^2\right]} \geq m,
\end{equation}
we have that either $\frac{F_Q(\rho,\hat{A}_{\text{Re}})}{\sum_{i=1}^N(\Delta a_{i,\text{Re}})^2}$ or $\frac{F_Q(\rho,\hat{A}_{\text{Re}})}{\sum_{i=1}^N(\Delta a_{i,\text{Im}})^2}$ equals or exceeds $m$ and so the wavefunction has at least $(m+1)$-partite entanglement by the standard arguments connecting the QFI of a Hermitian operator to the presence of multipartite entanglement.

\begin{figure}[htp!]
\centering
\includegraphics[width=0.5\linewidth]{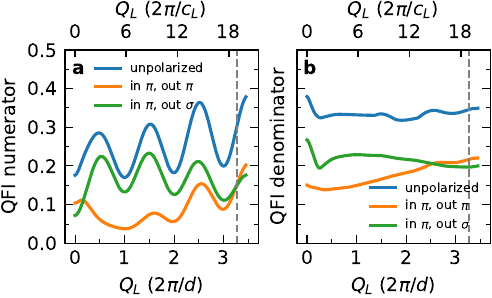}
\caption{\textbf{Unrenormalized QFI and maximal QFI for 1-partite states for the material Ba$_3$CeIr$_2$O$_9$ as a function of momentum transfer.} This figure provides the numerator and denominator of the normalized QFI found in Fig.~6 in the main text. \textbf{a} \textit{Numerator}:  We plot here the unrenormalized QFI, with final state polarization discrimination, for different momentum transfers at fixed incident energy $\omega_{\text{in}}=11.216$ keV corresponding to Fig. 6 in the main text. The grey dashed line corresponds to the momentum at which we plot the QFI for different incident photon energies (in Fig. 7 of the main text). Here the $\pi$ polarization is in the incident plane, while the $\sigma$ polarization is out of the incident plane. \textbf{b} \textit{Denominator}: Same setup as \textbf{a} but now plotting the maximal QFI for 1-partite states.} \label{fig:CeMomdepNonorm}
\end{figure}

\begin{figure}[htp!]
	\centering
	\includegraphics[width=0.5\linewidth]{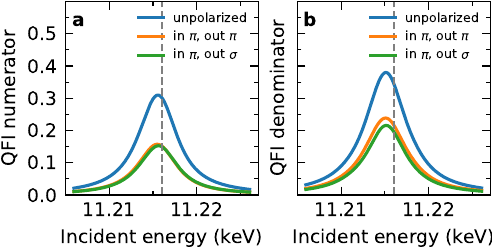}
\caption{\textbf{Unrenormalized QFI and maximal QFI for 1-partite states for the material Ba$_3$CeIr$_2$O$_9$ for Fig.~7 in the main text.} Same as Fig. \ref{fig:CeMomdepNonorm} above but now plotted as a function of incident energy, providing the numerator \textbf{a} and denominator \textbf{b} used to construct the normalized QFI found in Fig.~7 in the main text. For better comparison, the fixed energy used for the momentum-dependent scan on Fig. \ref{fig:CeMomdepNonorm} is indicated with a dashed gray line.}
\label{fig:CeEdepNonorm}
\end{figure}

\subsection{Determining the QFI bounds while taking into account twisted parity}
As we discuss in the main text, using the standard bounds of the QFI based on Eqn.~4 of the main text combined with Eqn.~\ref{QFIsum} above do not work as these bounds are predicated on operator locality, i.e.\ the individual $\hat A_i$ have to only act non-trivially on the site $i$.  However because of the presence of the denominator $1/(w_{\rm in}+i\Gamma_c-E_n+H_0)$ in the definition of the RIXS operator, $\hat A_R$, this is not true, absent the ultra short core hole lifetime approximation, an approximation we do not want to make here.  The other issue, also identified in the main text, with the bounds as described in Sec.~IV A above is that they do not take into account any of the symmetries of the theory.  They thus allow for the possibility of unphysical wavefunctions contributing to the bound.  It would be better to have bounds that reflect the natural symmetries present in the materials.  This has the advantage that the refined bounds will make it easier to detect entanglement.

To this end we restrict the wavefunctions over which we construct the bounds in two ways: i) we only consider wavefunctions with definite electron number and ii) for the Ce-iridate material, we consider wavefunctions that have definite twisted parity.
As we discuss in the main text for Ba$_3$CeIr$_2$O$_9$, there are 10 electrons in total shared among the two sites in the dimer of this material.  In our theoretical description of this material, we have limited ourselves to the excitations in the $t_{2g}$ manifold as our data set does not uniformly include spectral information that covers spectral weight produced from transitions into the $e_g$ manifold.  Thus the set of 1-partite states that respect parity over which we construct the bound will have 5 electrons on each iridate atom.  This leaves six possible states $\{|B_u\rangle\}^6_{j=1}$ for each site.  A generic wavefunction on either of the two sites is then given by $|\alpha\rangle=\sum^6_j a_j |B\rangle_j$.  A 1-partite wavefunction for the dimer that respects parity is then given by
$$
|\psi_\alpha\rangle=|\alpha\rangle\otimes|\alpha\rangle.
$$
It is then straightforward numerically to find the maximum of the QFI for such 1-partite states.
We show the results of the construction of these QFI bounds on the right hand side of Figs.~\ref{fig:CeMomdepNonorm} and \ref{fig:CeEdepNonorm}.  These bounds are those used in constructing the normalized QFI presented in Figs.~6 and 7 of the main text.

For the material, Ba$_3$TaIr$_2$O$_9$, there are 11 electrons shared between the two iridium ions.  Any 1-partite state that then respects electron number conservation will then have unequal numbers of electrons on the two iridate atoms.  Thus any state that respects both parity and number conservation will be 2-partite.  Thus here, unlike with Ba$_3$CeIr$_2$O$_9$, for the purposes of constructing the 1-partite bounds, we do not insist the wavefunction respect parity.  Furthermore for the Ta-material, unlike for the Ce-material, we have allowed states that had up to one $e_g$ electron.  In the Ce-case, the construction of the maximal QFI for 1-partite states involved 5 free complex parameters once wavefunction normalization was taken into account.  For the Ta-material, we maximized the 1-partite QFI separately in the $N_1=5$, $N_2=6$  and the $N_1=6$, $N_2=5$ sectors, each involving a distinct 5-parameter optimization.


\subsection{Maximum QFI for a 2-partite state}

In this section, we characterize the nature of the state that maximizes the QFI for both even and odd parity states in the cerium iridate.  Because of the relative complexity of the individual iridate sites, it is worthwhile to explore the nature of these states in greater detail.

We begin by constructing the definite-parity two-partite states for Ba$_3$CeIr$_2$O$_9$, focusing on how the nature of the state that maximizes the QFI evolves as a function of the momentum transfer encoded in the RIXS operator $\hat A_R$. In accordance with the model used, we work in a number-conserving two-site Fock space with a fixed total valence occupancy of $N_{\mathrm{tot}}=10$ electrons and with the additional constraint that no electron occupies the $e_g$ orbitals. This yields a restricted 66-dimensional basis consistent with $t_{2g}$-only configurations:
\begin{equation}
\dim \mathcal{H}_{\mathrm{val}} = \binom{12}{10} = 66
= \binom{6}{4}\binom{6}{6} + \binom{6}{5}\binom{6}{5} + \binom{6}{6}\binom{6}{4},
\end{equation}
corresponding to the three local occupancy sectors $(N_1,N_2)=(4,6)$, $(5,5)$, and $(6,4)$ where $N_i$ is the number of electrons sitting on site $i$.

We obtain the even-parity subspace by diagonalizing the (twisted) parity operator $P$ and collecting eigenvectors, $n_+$ in total, with eigenvalue $+1$. Denoting by $Q_+$ the $66\times n_+$ matrix whose columns form an orthonormal basis of the $+1$ eigenspace, any even-parity state is parameterized as
\begin{equation}
|\psi\rangle = Q_+\,|c\rangle, \qquad |c\rangle\in \mathbb{C}^{n_+}.
\end{equation}
Operators entering the optimization are restricted consistently via
\begin{equation}
O \mapsto O_+ = Q_+^\dagger O\,Q_+.
\end{equation}
For each momentum point (specified by the $L$ component in reciprocal lattice units) and for each polarization channel, we construct the relevant effective operator matrix in the many-body eigenbasis obtained from exact diagonalization. These matrices are then decomposed into Hermitian components as required by the objective functional.

We now numerically maximize the target functional over normalized pure states in the even-parity sector. The variational state coefficients $c\in\mathbb{C}^{n_+}$ are parameterized by nonnegative moduli and phases,
\begin{equation}
c_k = r_k e^{i\theta_k},\qquad r_k\ge 0,\qquad \sum_k r_k^2 = 1.
\end{equation}
The optimization is done over an unconstrained real vector $u$ and setting $r=u/\|u\|$.
At each value of momentum transfer, we perform multi-start local optimization using the L-BFGS-B algorithm with simple bounds on $r_k$ and $\theta_k\in[0,2\pi)$. An analytic gradient of the objective with respect to $(u,\theta)$ is supplied to the optimizer. Among the random restarts the best optimum is retained as the maximizer. For the odd-parity sector (eigenvalue $-1$) the method is entirely analogous.

In addition, we track the weight of the optimized state in selected local occupancy sectors (e.g.\ $(5,5)$ by summing probabilities over the corresponding index subsets in the product basis).

\begin{figure}[htp!]
        \centering
        \includegraphics[width=0.5\linewidth]{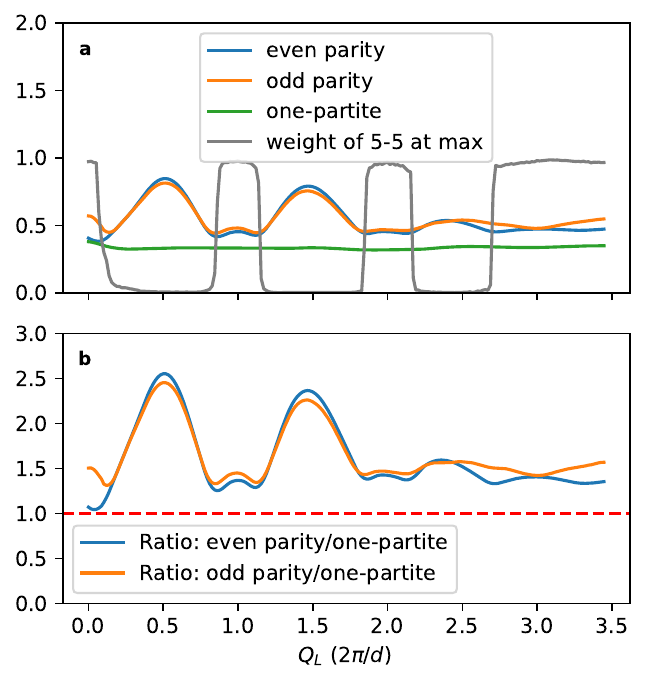}
\caption{\textbf{An analysis of the entanglement character of the states maximizing 1- and 2-partite QFI for the material Ba$_3$CeIr$_2$O$_9$.} We perform this analysis as a function of momentum transfer for fixed incident energy $\omega_{\text{in}}=11.216$ keV. \textbf{a}  In this panel we compare the maximum value of QFI for one-partite states (green) with generic even-parity states (blue) and odd-parity states (orange), as a function of momentum transfer. The gray line labeled 'weight of 5-5 at max', corresponds to the sum of the squares of the coefficients of the state maximizing the QFI belonging to states in the $(5,5)$ sector. \textbf{b} In this panel we plot the ratio of the maximal QFI in each sector with the one-partite maximum.}
        \label{fig:CeExpNonorm}
\end{figure}
Fig.~\ref{fig:CeExpNonorm} summarizes (i) the maximal value of the unrenormalized QFI within the even- and odd-parity sector as a function of momentum transfer as compared to the corresponding maximal one-partite QFI, and (ii) the maximal QFI normalized with respect to the one-partite QFI maximum. We find that the even-parity optimization is characterized by (at least) two distinct, competing families of local maxima. One family is dominated by the number-imbalanced sectors $(N_1,N_2)=(6,4)$ and $(4,6)$, i.e. by states well-approximated as (anti)symmetric superpositions of the form
$|\psi_{64}\rangle \propto |6\rangle\otimes|4\rangle \pm | 4\rangle\otimes|6\rangle$, whereas the other family is dominated by the balanced sector $(5,5)$.
Which of these solutions provides the global maximum depends sensitively on the momentum transfer, leading to an extended momentum dependence that is effectively ``binary-like'': the optimizer switches between a $(6,4)\oplus(4,6)$-dominated maximizer and a $(5,5)$-dominated maximizer, with little hybridization between these sectors over broad momentum ranges.

The entanglement structure of the $(6,4)\oplus(4,6)$-dominated maximizer is strongly constrained by parity and by the structure of the local Hilbert space. In particular, the local $N=6$ subspace is one-dimensional (a completely filled $t_{2g}^6$ configuration), so any state supported purely on $(6,4)\oplus(4,6)$ can be written as a sum of at most two product terms and therefore has Schmidt rank $\le 2$; enforcing even parity fixes the relevant linear combination to the symmetric form above. By contrast, the $(5,5)$-dominated maximizer can access a larger local manifold on each site, typically leading to a broader distribution of Schmidt weights. Consistent with this, while the $(6,4)\oplus(4,6)$ solution can yield a larger optimized witness (QFI) values at certain momenta, the entanglement entropy tends to be larger in the $(5,5)$-dominated regime,

\section{Model Parameter Inference with Bayesian Optimization Including Error Analysis}

We have adopted the method of \cite{Lajer2025Ham} to infer the needed Hamiltonian model parameters from experimental RIXS spectra. Following \cite{Lajer2025Ham}, we used a two-step optimization process consisting of a Gaussian Process Regression (GPR) step, followed by a greedy refinement of the most promising candidate points. The optimization involved a distance function based on the pixel-wise L1 distance $\chi^2_{L_1}$ between model-simulated (parameter-dependent) spectra and the experimental "ground truth". This normalized \( L_1 \) distance is defined as
\begin{equation}
  \chi^2_{L_1} = \sum_{ij} \left| r_{ij} - s_{ij} \right|,
\end{equation}
where 
  \[
    r_{ij}=\frac{R_{ij}}{\sum_{kl} R_{kl}}\quad s_{ij}=\frac{S_{ij}}{\sum_{kl} S_{kl}}. 
  \]
In the above formula, we denote the entries of the 2D arrays containing the experimental (reference) and simulated spectral intensities by $R_{ij}$ and $S_{ij}$, respectively. The summation over indices $i,j,k,l$, etc. spans over each row and column of the corresponding arrays.

In the case of $\text{Ba}_3\text{CeIr}_2\text{O}_9$, we have two primary RIXS data sets: i) a data set at fixed momentum transfer $Q=(-0.5,0,18.94)$ where incident photon energy is swept (as pictured in Fig. 2 of the main text) and ii) a data set at fixed incident photon energy (11.216 keV) where momentum transfer is swept.  To infer the Hamiltonian parameters from these two sets, we focus on the energy loss interval $0.2~\mathrm{eV} < E_{\text{loss}} < 2~\mathrm{eV}$.  This energy loss interval is available in both data sets.  This restricted availability is partly why we focused on modelling only the intra $t_{2g}$ excitations for the Ce-material whose spectral contributions are found below 2 eV.  Distances for the energy and momentum dependent spectra were combined by simple addition. 

For $\text{Ba}_3\text{TaIr}_2\text{O}_9$, we again have two data sets, one at fixed momentum transfer with incident photon energy swept (Fig. \ref{fig:TaEmap})  and one at fixed incident photon energy with momentum transfer swept (Fig. \ref{fig:TaLmap}).  Because this material has 11 electrons (as opposed to the 10 its cerium counterpart), we needed to expand the Hilbert space of the model to range beyond the $t_{2g}$ manifold, now allowing up to one electron to be present in an $e_g$ state.  We correspondingly set the energy loss interval to 0.3eV$<E_{\text{loss}}<2$eV for the spectra with momentum transfer dependence (Fig. \ref{fig:TaLmap}) (we could not go beyond 2eV as we do not have data at higher energy losses) and 0.3eV$<E_{\text{loss}}<6$eV for the spectra with incident photon energy dependence (Fig. \ref{fig:TaEmap}). For the Ta-material as with the Ce-material, we omitted the region of small $|E_{\rm loss}|$ in our parameter inference procedure to remove non-physical artifacts from the broadened elastic peak. 

In the GPR step, we performed $54$ independent runs, each consisting of $1000$ iterations,  for both materials. In this first step, we floated the hopping parameters $V_{dd,\sigma}$, $V_{dd,\pi}$, $V_{dd,\delta}$, the Slater parameters $F^2_{dd}$, $F^4_{dd}$, $F^2_{dp}$, $G^1_{dp}$, $G^3_{dp}$ and spin-orbit couplings $\zeta$ and $\zeta_c$. For the Ce complex, the GPR analysis resulted in $22$ candidate points with $\chi^2_{L_1} \leq 1.2\chi^2_{L_1,min}$, where $\chi^2_{L_1,min}$ is the smallest distance found. For the Ta material, the GPR procedure yielded $24$ candidate points with $\chi^2_{L_1} \leq 1.15\chi^2_{L_1,min}$.

These parameter sets were subsequently refined further in the greedy optimization step. The refinement relied on the PY-BOBYQA algorithm, maximizing the number of function evaluations in $3600$ evaluations per point. The greedy refinement allowed the increase of floated parameters as well: besides the GPR parameters, we also tuned $10Dq$ and the core-hole broadening $\Gamma_c$. The parameters $\kappa$, $\theta$, $a_0$, $U_{dd}$ and $U_q$ as well as the final state broadening $\Gamma_f$ were kept fixed during the fitting process.  

For the Ce material, a number of parameter sets saturated the fitting bounds resulting in an essentially zero predicted value for $F^2_{dd}$. We excluded these solutions on physical grounds.  This left us with a set of parameter configurations which led to a reasonable replication of the experimental data for the Ce-material.  We used the parameter set with the overall minimal L1 distance $\chi^2_{L_1}$ found for producing the plots in Figs. 6 and 7 in the main text. Apart from the overall best solution, we also computed the QFI for a number of other candidate solutions (with competitive $\chi^2_{L_1}$ values).  The plotted QFI values are the averages over these configurations while the error bars are the standard deviations.

In the case of Ta material, our parameter inference identified a set of 24 candidate parameter configurations. Unlike with the Ce-material, we did not filter our results, as these were not contaminated by unphysical parameter sets.  The plots in Figs. \ref{fig:TaEmap} and \ref{fig:TaLmap} the parameter set with the smallest L1 distance.  To obtain an uncertainty for the QFI for the Ta-material, we computed the QFI for each of the 24 configurations and plotted the averages and standard deviations in Figs. \ref{fig:TaLQFI} and \ref{fig:TaEQFI}. 

\begin{table*}
\caption{Full list of parameters used in the exact diagonalization calculations for Ba$_3$TaIr$_2$O$_9$. As mentioned in the main text, the crystal field splitting parameters such as $\theta$ and $\kappa$ are defined in Ref.~\onlinecite{Kugel2015Spin}, and $\zeta$ is the spin-orbit coupling parameter for the Ir $5d$ orbitals while the one for the $2p$ core orbitals is fitted to be 1144.174~eV. The inverse core-hole lifetime is fitted to be 2.73~eV. All parameters except $\theta$ and $\kappa$ are in units of eV.}
\begin{ruledtabular}
\begin{tabular}{cccccccc}
\multicolumn{4}{c}{Crystal field splitting and spin-orbit coupling} & & \multicolumn{3}{c}{Hopping integrals}\\
$10D_q$ & $\theta$ & $\kappa$ & $\zeta$ & & $V_{dd\sigma}$ & $V_{dd\pi}$ & $V_{dd\delta}$\\
3.28 & 49.00 & 0.1 & 0.33 & & -0.95 & -0.53 & 0.23\\
\hline
\multicolumn{4}{c}{Core-hole potential} & & \multicolumn{3}{c}{On-site Coulomb interactions}\\
$F^0_{dp}$ & $F^2_{dp}$ & $G^1_{dp}$ & $G^3_{dp}$ & & $F^0_{dd}$ & $F^2_{dd}$ & $F^4_{dd}$\\ 
2.506 & 0.0 & 0.001 & 0.155 & & 2.731 & 2.409 & 0.880\\

\end{tabular}
\end{ruledtabular}
\label{table:allparamsTa}
\end{table*}

\section{Detection of Quantum Entanglement in $\rm Ba_3TaIr_2O_9$.}

\begin{figure}[htp!]
	\centering
	\includegraphics[width=0.5\linewidth]{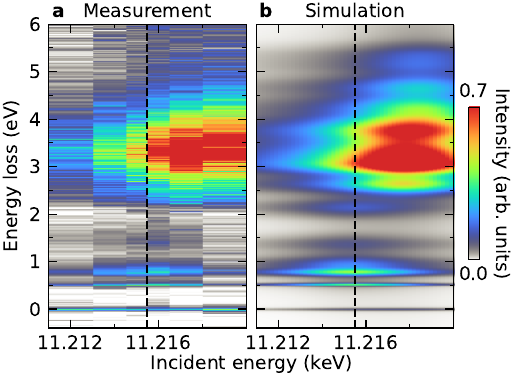}
	\caption{\textbf{Comparison of the incident energy dependence of the RIXS spectra between measurement and simulation for Ba$_3$TaIr$_2$O$_9$} {\bf a} Measured RIXS spectra with varying incident energy at fixed momentum transfer with $L=18.4$ in reciprocal lattice units (r.l.u.). The intensity below 1.5~eV energy loss correspond to intra-$t_{2g}$ transitions while that above 1.5~eV mainly comes from inter-$t_{2g}$-$e_g$ transitions. \textbf{b} Calculated incident-energy-dependent RIXS spectra at the same fixed momentum transfer. In both panels, the dashed line indicates the incident energy used for the momentum scan on Fig. \ref{fig:TaLmap}.}
	\label{fig:TaEmap}
\end{figure}

\begin{figure}[htp!]
	\centering
	\includegraphics[width=0.5\linewidth]{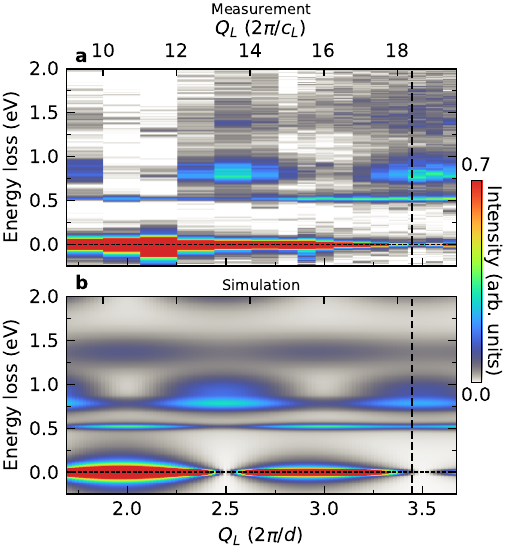}
	\caption{\textbf{Comparison of the $L$ dependence of the RIXS spectra between measurement and simulation for Ba$_3$TaIr$_2$O$_9$} {\bf a} Measured $L$-dependent RIXS spectra showing periodic modulation with the incident energy $\hbar\omega_{\text{in}}$ fixed to 11.216~keV. Only the intra-$t_{2g}$ transitions are presented since they are the dominating signals at this particular incident photon energy. Here $Q_L$ is the momentum transfer along the $L$ direction. For convenience, the same momentum scale is displayed in two units. Along the top axis, we use units of $2\pi/c_L$ where $c_L$ is the unit cell lattice constant along the $L$ direction. On bottom axis, we use units of $2\pi/d$, where $d$ is the distance between the two dimers along the $L$ direction. \textbf{b} Calculated $L$-dependent RIXS spectra at the same fixed incident energy. In both panels, the dashed line indicates the momentum used for the incident energy scan on Fig. \ref{fig:TaEmap}.}
	\label{fig:TaLmap}
\end{figure}

\begin{figure}[htp!]
	\centering
	\includegraphics[width=0.5\linewidth]{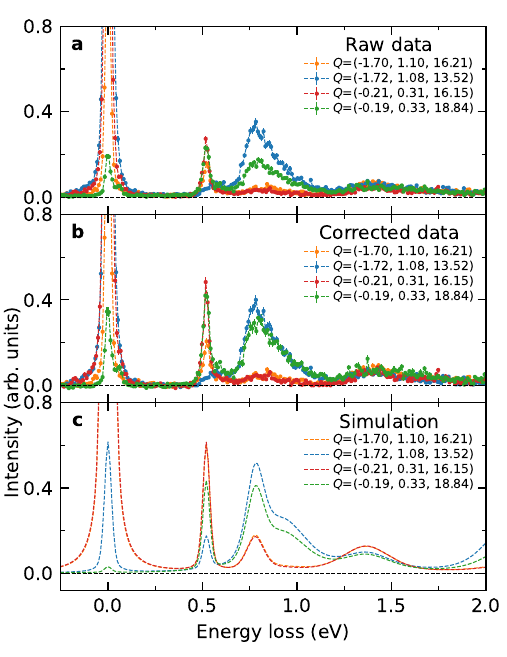}
	\caption{\textbf{Comparison of RIXS spectra between measurement and simulation at representative momentum transfers for Ba$_3$TaIr$_2$O$_9$.} \textbf{a} Representative RIXS spectra at the indicated momentum transfers in r.l.u., and the incident energy $\hbar\omega_{\text{in}}$ is fixed to 11.216~keV. Here the raw data is shown before applying standard corrections. Only the intra-$t_{2g}$ transitions are presented since they are the dominating signals at this particular incident energy. \textbf{b} Same as \textbf{a} after self-absorption correction and background removal. \textbf{c} Calculated RIXS spectra at the same momentum transfers and the same fixed incident energy.}
	\label{fig:TaOmap}
\end{figure}

\begin{figure}[htp!]
	\centering
	\includegraphics[width=0.5\linewidth]{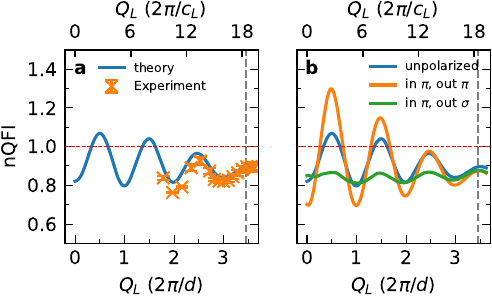}
\caption{\textbf{The dependence of the QFI for  for Ba$_3$TaIr$_2$O$_9$ as a function of momentum transfer.} \textbf{a} Comparison of the QFI as determined by the simulation and measurement, for different momentum transfers at fixed incident energy $\hbar\omega_{\text{in}}=11.216$ keV. \textbf{b} Same setup as \textbf{a} but with final state polarization discrimination. The legend in the right panel provides the QFI for when we resolve the polarization of the final state, where the $\pi$ polarization is in the incident plane, $\sigma$ polarization is out of the incident plane. The red dashed line on both panels is the nQFI threshold for 2-partite entanglement. The dashed gray line marks the momentum transfer at which the incident photon energy is swept in Fig.~\ref{fig:TaEQFI}.}
	\label{fig:TaLQFI}
\end{figure}

\begin{figure}[htp!]
	\centering
	\includegraphics[width=0.5\linewidth]{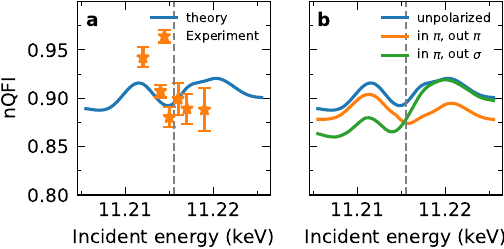}
	\caption{ {\bf The dependence of the QFI for Ba$_3$TaIr$_2$O$_9$ as a function of incident photon energy.}  \textbf{a} comparison of simulation and measurement, for different incident photon energies at fixed momentum transfer $L=18.4$ r.l.u. \textbf{b} Same setup as \textbf{a} but with final state polarization discrimination. The legend in the right panel documents the polarizations for the final state, where the $\pi$ polarization is in the incident plane, $\sigma$ polarization is out of the incident plane. The dashed gray line marks the incident photon energy at which the momentum transfer is swept in Fig.~\ref{fig:TaLQFI}.}
	\label{fig:TaEQFI}
\end{figure}

We have performed the same entanglement analysis for the isostructural material, Ba$_3$TaIr$_2$O$_9$, as we did for Ba$_3$CeIr$_2$O$_9$. In this material, however, 11 electrons are shared between the two Ir ions instead of the 10 in Ba$_3$CeIr$_2$O$_9$, leading to distinct RIXS features. We follow the same strategy as the main text and use the codebase EDRIXS \cite{WANG2019151} with tuned parameters in Table~\ref{table:allparamsTa} below to reproduce the experimentally measured RIXS intensity and construct the corresponding RIXS operator at the same time. To mimic the observed lineshape of the excitation peaks, the final-state energy loss spectra are broadened using multiple pseudo-Voigt profiles with the fraction fixed to 0.5 and full-widths at half-maximum of 0.044 and 0.108~eV for the first and second excitation peaks, respectively, and 0.3~eV for the remaining. The comparisons between the measurement and simulation are shown in Figs.~\ref{fig:TaEmap}, \ref{fig:TaLmap}, and \ref{fig:TaOmap} here, exhibiting excellent agreement between the model and the measured spectra. Using the constructed RIXS operator, the nQFI for Ba$_3$TaIr$_2$O$_9$ was subsequently calculated for different incident energies and momentum transfers, and the results are shown in Figs.~\ref{fig:TaLQFI} and \ref{fig:TaEQFI} (analogous to Figs. 6 and 7 in the main text). We see in Fig.~\ref{fig:TaLQFI} that the QFI hovers around 1 for the entire range of $Q_L$. 2-partite entanglement is detected in terms of our modeling for smaller values of $Q_L$, although not in the region where we have experimental data.  We do see however that the experimental values of the QFI closely track the modeled values and so we would expect experimental measurements at small $Q_L$ would in fact detect bipartite entanglement.  In Fig.~\ref{fig:TaEQFI} we consider the QFI across a range of incident photon energies.  Here the QFI hovers around 0.9 so missing detecting bi-partite entanglement.  Here as well the experimental values of the QFI are in good agreement with the modeled values.  We also consider the effects of polarization discrimination upon the values of the QFI in Figs.~\ref{fig:TaLQFI} and \ref{fig:TaEQFI}.  We see that with polarization discrimination, we find higher values of the QFI for $\pi$-polarizations of the incoming and outgoing photons at small momentum transfer (see Fig.~\ref{fig:TaLQFI}).  We however see a much weaker sensitivity to polarization selection as a function of incident photon energy -- see Fig.~\ref{fig:TaEQFI}.

\bibliography{main}